\documentclass[preprint,3p,sort&compress]{elsarticle}
\journal{Elsevier}
\usepackage[usenames,dvipsnames]{xcolor}
\usepackage{mathptmx}
\usepackage{graphicx}           % standard LaTeX graphics tool
\usepackage{amsmath}
\usepackage{amsfonts}
\usepackage{amssymb}
\usepackage{epstopdf}
\usepackage{graphicx}
\usepackage{tikz}
\usepackage{xspace}
\usepackage{flowchart}
\usepackage{color}
\usepackage{subcaption}
\usepackage{booktabs}
\usepackage{bm}
\usepackage{algorithm}
\usepackage{algpseudocode}
\newlength\myindent
\newcommand{\nc}{\newcommand}
\nc{\rnc}{\renewcommand}
\nc{\bs}{\boldsymbol}
\nc{\mrm}{\mathrm}
\nc{\mum}{$µ$\mrm{m}}
\rnc{\matrix}[2]{\left[\!\!\begin{array}{#1}
	#2\end{array}\!\!\right]}
\rnc{\vector}[1]{\matrix{c}{#1}}
\nc{\mm}[1]{\boldsymbol{#1}}
\nc{\mms}[1]{\boldsymbol{#1}}
\nc{\real}[1]{\mathrm{Re} \lbrace #1 \rbrace}
\nc{\imag}[1]{\mathrm{Im} \lbrace #1 \rbrace}
\nc{\dd}{\mathrm{d}}
\nc{\ii}{\mathrm{i}}
\nc{\ee}{\mathrm{e}}
\nc{\inv}{^{-1}} %not used
\nc{\herm}{^{\mathrm H}}
\nc{\tra}{^{\mathrm T}}
\nc{\conj}[1]{ \overline{#1} }
\nc{\normal}{\mathrm n}
\nc{\tangential}{\mathrm t}
\nc{\kn}{{k_{\normal}}}
\nc{\kt}{{k_{\tangential}}}
\nc{\ie}{i.\,e.\xspace}
\nc{\eg}{e.\,g.\xspace}
\nc{\cf}{cf.\xspace}
\nc{\myquote}[1]{`#1'}
\nc{\etal}{et al.\xspace}
\nc{\fabstand}{\,}
\nc{\fp}{\fabstand.}
\nc{\fk}{\fabstand,}
\nc{\x}[1]{\mbox{#1}}
\nc{\kbb}{\mm K_{\mrm{bb}}}
\nc{\kbi}{\mm K_{\mrm{bi}}}
\nc{\kib}{\mm K_{\mrm{bi}}^{\mrm{T}}}
\nc{\kbbinv}{\mm K_{\mrm{bb}}^{-1}}
\nc{\kii}{\mm K_{\mrm{ii}}}
\nc{\dii}{\mm D_{\mrm{ii}}}
\nc{\mii}{\mm M_{\mrm{ii}}}
\nc{\mbb}{\mm M_{\mrm{bb}}}
\nc{\db}{\mm d_{\mrm{b}}}
\nc{\ai}{\mm a_{\mrm{i}}}
\nc{\vi}{\mm v_{\mrm{i}}}
\nc{\di}{\mm d_{\mrm{i}}}
\nc{\dvi}{\dot{\mm v}_{\mrm{i}}}
\nc{\ddi}{\dot{\mm d}_{\mrm{i}}}
\nc{\fex}{\mm f_{\mrm{ex}}}
\nc{\fexb}{\mm f_{\mrm{ex,b}}}
\nc{\fexi}{\mm f_{\mrm{ex,i}}}
\nc{\wb}{\mm W_{\mrm{b}}}
\nc{\wbtra}{\mm W_{\mrm{b}}^{\mrm T}}
\nc{\g}{\mm g}
\nc{\h}{\mm h}
\nc{\gam}{\mm \gamma}
\nc{\lam}{\mm \lambda}
\nc{\C}{\mm C}
\nc{\nC}{C}
\nc{\dt}{\Delta t}
\nc{\dg}{\Delta \g}
\rnc{\active}{\mrm{a}}
\nc{\projC}[1]{{\operatorname{proj}}_{\mathcal C}\left(#1\right)}
\nc{\projCA}[1]{{\operatorname{proj}}_{\mathcal C_{\active}}\left(#1\right)}
\nc{\qmod}{q_{\mrm{mod}}}
\nc{\Dmod}{\zeta}
\nc{\ABAQUS}{\textsf{ABAQUS}\xspace}
\nc{\SIERRA}{\textsf{SIERRA}\xspace}
\nc{\MATLAB}{\textsf{MATLAB}\xspace}
\nc{\MSM}{\textsf{MSM}\xspace}
\nc{\NLvib}{\textsf{NLvib}\xspace}
\nc{\NLstep}{\textsf{NLstep}\xspace}
\nc{\MBCMS}{\textsf{MB-CMS}\xspace}
\nc{\epsAL}{{\mm\varepsilon}_{\mrm{AL}}}
\nc{\fig}[4][tbh]{
\begin{figure}[#1]
\centering
\includegraphics[width=#4\textwidth]{figures/#2}
\caption{#3\label{fig:#2}}
\end{figure}}
\nc{\e}[2]{\begin{equation} #1 \label {eq:#2} \end{equation}}
\nc{\est}[1]{\begin{equation*} #1 \end{equation*}}
\nc{\ea}[1]{
\begin{eqnarray}
#1 \end{eqnarray}}
\nc{\east}[1]{
\begin{eqnarray*}
#1 \end{eqnarray*}}
\nc{\fref}[1]{{Fig.~\ref{fig:#1}}}
\nc{\frefo}[1]{{\ref{fig:#1}}}
\nc{\frefs}[1]{{Figs.~\ref{fig:#1}}}
\nc{\tref}[1]{{Tab.~\ref{tab:#1}}}
\nc{\trefo}[1]{{\ref{tab:#1}}}
\nc{\trefs}[1]{{Tab.~\ref{tab:#1}}}
\nc{\eref}[1]{{Eq.~(\ref{eq:#1})}}
\nc{\erefo}[1]{(\ref{eq:#1})}
\nc{\erefs}[1]{{Eqs.~(\ref{eq:#1})}}
\nc{\erefFull}[1]{Equation~(\ref{eq:#1})}
\nc{\sref}[1]{{Section~\ref{sec:#1}}}
\nc{\srefo}[1]{\ref{sec:#1}}
\nc{\srefs}[1]{{Sections~\ref{sec:#1}}}
\nc{\ssref}[1]{{Section~\ref{sec:#1}}}
\nc{\ssrefo}[1]{\ref{sec:#1}}
\nc{\ssrefs}[1]{{Sections~\ref{sec:#1}}}
\nc{\aref}[1]{{{\ref{asec:#1}}}}
\nc{\arefo}[1]{{\ref{asec:#1}}}
\nc{\arefs}[1]{{{Appendices~\ref{asec:#1}}}}

\makeindex

\begin{document}

\begin{frontmatter}
\title{
Enabling topography-resolving structural dynamic contact simulation
}
\author{
Hendrik D. Linder$^1$,
David A. Najera-Flores$^2$,
Robert J. Kuether$^3$,
Malte Krack$^1$,
\\
}
\address{$^1$ University of Stuttgart, GERMANY}
\address{$^2$ ATA Engineering, Inc., USA}
\address{$^3$ Sandia National Laboratories, USA}

\begin{abstract}
Damping of structures and systems is often dominated by frictional dissipation in connections, the prediction of which remains a longstanding scientific challenge.
Previous studies have shown that the actual topography of contact interfaces may have a strong effect, especially in the partial slip/liftoff regime.
We recently proposed a multi-scale method,
which couples finite element and boundary element modeling.
The primary benefit of this approach is to analyze the effect of actual contact topography on the dynamics of jointed structures.
While this multi-scale modeling method was initially developed for quasi-static analysis, we demonstrate herein how it can be used for time step integration and Harmonic Balance analysis.
We cross-verify those fully dynamic analysis methods against each other and quasi-static results, for the S4 Beam benchmark.
We compare the multi-scale method against state-of-the-art full-FE analysis, in terms of numerical damping and computational performance.
Some discrepancy is found to be of physical origin.
Depending on the load history, it is shown that the system settles to a slightly different equilibrium.
Transient multi-scale simulations enable the prediction of this interesting phenomenon, for the first time, for a structure with bolted joints.
\end{abstract}


\end{frontmatter}

\section{Introduction\label{sec:intro}}
% FRICTIONAL DAMPING IS OF TECHNICAL RELEVANCE
To engineer structures and systems against undesired noise and harmful vibrations, it is crucial to quantify their damping. % [aerospace|mechanical|civil]
The amount of \emph{damping} determines, in particular, if resonances can be survived, how prone a system is to dynamic instabilities, and how quickly cyclic stresses decay after a shock-type load.
In the absence of specific damping devices, \emph{dry friction} in mechanical connections, \eg, bolted or riveted joints, is the dominant cause of structural dissipation of most engineering systems \cite{Gaul.1997,Popp.2003}.
While monolithic design may offer weight reduction in some cases, jointed design prevails as it permits to combine different materials, facilitates manufacturing and maintenance/repair, and may lead to less waste.
The primary purpose of the described connections is to transmit load, provide stiffness, ensure leak tightness and/or alignment.
The contact interfaces are, therefore, designed to remain in a macroscopic stick phase under reasonable vibratory loads, while microscopic relative motion is allowed.
This regime is called \emph{partial slip}, pre-sliding or microslip.
Likewise, partial liftoff occurs, \ie, local opening-closing events, which also has a crucial effect on the structural dynamics.

\subsection{Partial slip/liftoff, and the importance of the actual topography}
% WE CANNOT PREDICT PARTIAL SLIP DAMPING YET; TOPOGRAPHY IS IMPORTANT
Over the past two decades, substantial research effort has been directed towards understanding and predicting friction damping.
Still, under partial slip/liftoff conditions, the predictive capabilities of today's state-of-the-art methods are quite limited \cite{Brake.2021}.
It is known that the actual contact interface topography may have a crucial impact on the initial pressure and gap distribution upon assembly, as well as the frictional-unilateral contact interactions under dynamic loading.
Real, manufactured surfaces feature form deviations, waviness and roughness. % DIN4760; \citeLiteratur{Yastrebov.2015}
The initial topography results from manufacturing and evolves during the system's lifetime for a number of reasons, including wear and corrosion.
In addition, misalignment affects the composite surface, \ie, the two-dimensional distance profile of the paired contact interfaces.
An important working hypothesis has been established over the past decade:
If the \emph{actual interface topography} is accounted for, reasonably accurate predictions of the vibration behavior can be obtained. % when imposing Coulomb-Signorini conditions; w.r.t frequency and damping ratio vs. amplitude of few lowest-order fundamental modes
This is supported, among others, by the theoretical and experimental works \cite{Willner.2004,Pesaresi.2017,Armand.2018,Brink.2020,Pinto.2023,ZareEstakhraji.2023,Porter.2023,Yuan.2023}.
Although there seems to be consensus on this hypothesis, the scales of topography that must be accounted for remains an open question.
Two fundamentally different approaches to account for the topography should be distinguished, which we will refer to as \emph{topography-resolving} and \emph{topography-blurring}, respectively.
\\
Topography-resolving simulations rely on a geometric model of the contact interfaces.
To obtain this, it is now common practice, at least in academic and component testing, to scan the interface topography using, \eg, coordinate measurement machines \cite{Wall.2022} or white-light interferometers \cite{Brink.2020,Fochler.2025,Goerke.2010}.
Provided that all relevant length scales are captured in this geometric model, it seems appropriate to model frictional-unilateral interactions among the surfaces via set-valued (rigid) contact laws (\emph{Coulomb-Signorini conditions}).
\\
In contrast to topography-resolving simulations, topography-blurring simulations use surface geometry data to inform compliant constitutive contact laws (if they are not entirely calibrated to tests) \cite{Balaji.2020}.
A combination is also possible, where the longer wave lengths of the topography are resolved, whereas the shorter ones are blurred.
Exemplary results of a topography-resolving simulation are shown in \fref{Rob} together with measurement data, in terms of the amplitude-dependent damping ratio of the first in-phase bending mode of a realization of the S4 Beam benchmark, which is also considered in the present work.
\begin{figure}[t]
  \centering
  \includegraphics[width=1.0\linewidth]{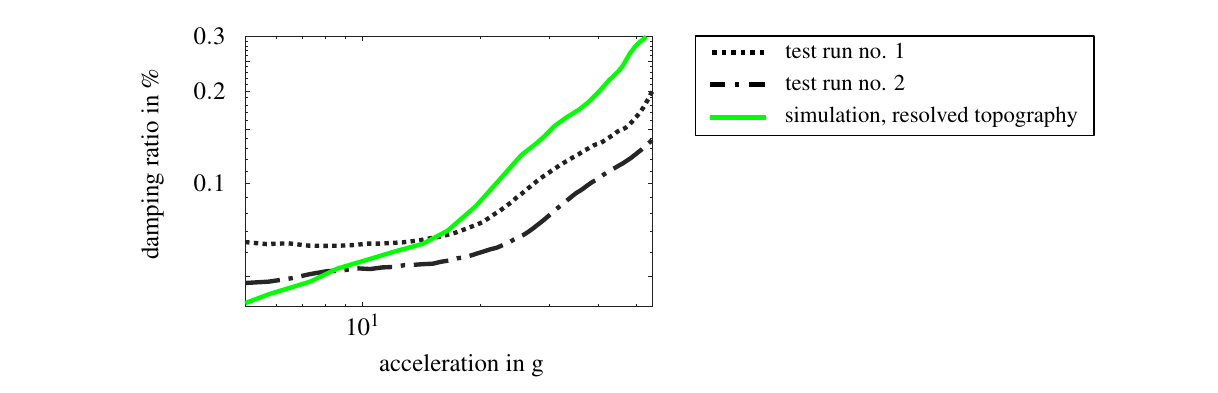}
 \caption{Experimental results and simulation results of a topography-resolving simulation of the S4 Beam.
 }
 \label{fig:Rob}
\end{figure}

\subsection{Computational challenges of topography-resolving simulations}
% COMPUTATIONAL CHALLENGES INVOLVED; MULTI-SCALE CHARACTER OF THE PROBLEM
Dynamic topography-resolving simulation is computationally demanding or even infeasible. % for the time window of interest, sufficiently invariant
An important reason for this is the multi-scale character of the problem:
The local relative displacements within a mechanical connection may be in the (sub-)micrometer range, while the maximum (absolute) vibration level of the system may be much larger, \eg, in the order of several millimeters.
Similarly, the form deviations may be in the micrometer range, while the dimensions of the nominal contact area may be in the range of centimeters, and the dimensions of the parts may be in the meter range.
This leads to extremely high model orders (\eg high numbers of nodes in FE models).
The second aspect that challenges computational feasibility is the set-valued contact laws (Coulomb-Signorini conditions).
These idealized contact laws are beneficial with regard to the aim for predictive modeling:
The friction coefficient remains as the only parameter of the contact model.
The current state of the art is to identify the friction coefficient from dedicated tests, and the transfer of this parameter to different assemblies seems justified, provided that the considered material pairing and ambient conditions are equivalent.
It should be emphasized that this is in complete contrast to contact models whose stiffness or hysteresis parameters are calibrated to tests, such as the popular 4-parameter-Iwan model \cite{Segalman.2005}, for which it is known that such a transfer is invalid.
However, the set-valued character of the force laws, in conjunction with the fine contact interface resolution is highly challenging for computations.
For this type of problem (finely resolved contact interface in combination with stiff or rigid contact laws), FE analysis is prone to numerical error and non-convergence, see \eg \cite{Jewell.2020}.

\subsection{State of the art of topography-resolving simulations}
% AVAILABLE METHODS: QSMA
\emph{Quasi-Static Modal Analysis} (QSMA) is the prevailing method for computing amplitude-dependent modal parameters of topography-resolving models.
The idea of that method is to estimate natural frequency and damping ratio of a jointed structure by quasi-statically loading the structure with a modal body force, and reconstructing the hysteresis curve using Masing rules \cite{Allen.2016,Lacayo.2019a}.
%The idea of that method is as follows:
%First, the static equilibrium configuration is obtained accounting for the assembly process (\eg tightening sequence of the bolts).
%Second, a conventional modal analysis step is done by assuming linear behavior around this configuration.
%Third, the static response to a load of the form, mass matrix times modal deflection shape, is applied, while the scale factor is successively increased.
%From the results, an initial loading curve can be computed in the modal force-displacement diagram.
%From this curve, a hysteresis cycle can be reconstructed for each computed displacement level using the Masing rules.
%From those results, finally, one obtains an estimate of the natural frequency and the damping ratio, of the considered mode, as function of the displacement level \cite{Allen.2016,Lacayo.2019a}.
% BENEFITS AND LIMITATIONS
An important benefit of QSMA is that all demanding analysis steps can be done with a conventional FE tool.
QSMA is a widespread method for jointed structures, and its application is not restricted to topography-resolving models.
The maturity of the method has been demonstrated in real-world applications, including the Orion Multi-Purpose Crew Vehicle \cite{Allen.2020}.
An important deficiency of QSMA is that it is limited to well-separated modal frequencies, weakly nonlinear behavior (\eg only a few percent modal frequency shift), and scenarios where the (global) modal deflection shape does not change significantly.
Locally at the contact interface, the deflection shape will change substantially with the displacement, as initially sticking parts of the interface start sliding or separating, and/or initially open parts come into contact.
Such contact transitions are allowed in QSMA.
Still the deviation between the actual deflection shape and that used in the applied forcing gives rise to some inconsistency, and limits the accuracy of that method.
In addition, since the common implementation of QSMA relies on the Masing rules, which are restricted to fixed normal load, inaccurate results are obtained in the case of substantial normal load variation \cite{Najera-Flores.2023}.
\\
% AVAILABLE METHODS: DYNAMICS
If the conditions of QSMA (well-separated modal frequencies, invariant deflection shape, weakly nonlinear behavior, approximately constant normal load) do not hold, one has to resort to (fully) dynamic analyses.
For the purpose of computing amplitude-dependent modal parameters, noteworthy alternatives, which overcome the deficiencies of QSMA, are modal analysis in accordance with the Extended Periodic Motion Concept \cite{Krack.2015}, typically carried out with the Harmonic Balance method, and the identification of modal parameters form the transient ring-down, \ie, the free decay following an appropriate impulsive load \cite{Kuether.2016}.
These techniques are employed in the present work and described in detail later.
Besides modal analysis, dynamic analyses are indispensable for transient dynamics, such as the response to seismic loads or explosions, or for steady-state dynamics, such as the response to periodic loads.
Numerical performance and results depend crucially on the algorithm for contact enforcement, the time step integration scheme, and the model order.
In the dynamic case, non-convergence is an even more severe problem than in the quasi-static case.
Even if the energy remains finite and the solver converges at each time step, numerical errors may severely distort the simulation results.
When the (usually very light) physical damping is a quantity of interest, numerical damping is a critical error type.
Although the total energy may be well-conserved, artificial energy transfer from low- to high-frequency modes can occur \cite{MonjarazTec.2022}.
This manifestation of numerical error is greatly under-appreciated in the view of the authors.
% NOTE: useful to emphasize unconditional stability; energy conservation properties etc. do not hold in case of contact (algorithm parameters tuned to strongly simplified scenarios like friction-less impact of 2 point masses)
Compared to quasi-static analyses, dynamic analyses of structures with finely resolved contact interfaces are rare.
Simulations of that kind, which we are aware of, are listed in \tref{dynamic_simulations}.
Most of these were reported in the past 5 years.
In particular, simulations involving $>2000$ nodes per interface, which would be needed for a reasonable resolution of the topography, are almost non-existent, and most of studies use rather soft penalty springs (without good physical/experimental justification).
\begin{table}[ht]
  \centering
  \caption{Structural dynamic simulations involving 3D contact in bolted joints or at friction dampers. BRB: Brake-Reuss Beam; UPD: under-platform damper; HPT: high-pressure turbine; HB: Harmonic Balance
  }
  \label{tab:dynamic_simulations}
  \vspace{0.3em}
  \begin{tabular}{ccccccc}
    \toprule
    benchmark & topology & interface & no. of nodes & contact stiffness & algorithm & reference\\
     &  & reduction & in contact & in $\frac{\mrm{kN}}{\mrm{mm^3}}$ &  & \\

    \midrule
    TRChallenge & flat & none & 51005 & $\infty$ (rigid) & explicit & \cite{Krack.2025} \\
    S4 Beam & flat & none & 2$\times$610 & $k_{\mrm n} = 378$ & HB & \cite{Kuether.2024} \\
            &      &      &              & $k_{\mrm t} = 189$  &         &     \\
    S4 Beam & flat & none & 2$\times$1037 & $k_{\mrm n} =341$ & explicit & \cite{Hughes.2021} \\
            &      &      &              & $k_{\mrm t} = 225$  &         &     \\
    S4 Beam & flat & 200 modes & 2$\times$3350 & $k_{\mrm n} = 10$ & implicit & \cite{Pichler.2021} \\
            &      &      &              & $k_{\mrm t} = 30$  &         &     \\
    S4 Beam & flat & 139 modes & 2$\times$1677 & $k_{\mrm n} = 10$ & implicit & \cite{Witteveen.2023} \\
            &      &      &              & $k_{\mrm t} = 10$  &         &     \\
    BRB & flat & none & 592 with 86 active & $k_{\mrm n} = 6$ & HB & \cite{Lacayo.2019} \\
            &      &      &              & $k_{\mrm t} = 6$  &         &     \\
    BRB & flat & none & 592 & $k_{\mrm t} = 0.52$ & HB & \cite{Gross.2016} \\
    HPT UPD & flat & none & 2098 & $k_{\mrm n} = 37.5$ & HB & \cite{Fantetti.2023} \\
            &      &      &              & $k_{\mrm t} = 30$  &         &     \\
    cantilever beam & worn & none & 169 & $k_{\mrm n} = 246 $ & HB & \cite{Tamatam.2023} \\
            &      &      &              & $k_{\mrm t} = 198$  &         &     \\
    \bottomrule
\end{tabular}
\end{table}

\subsection{The recently proposed multi-scale method\label{sec:introMSM}}
% FE-BE MULTI-SCALE METHOD
We recently proposed a \emph{multi-scale method (\MSM)}, where the structural dynamics is described with a relatively coarse FE model, and the contact mechanics with a sufficiently fine Boundary Element (BE) model \cite{Linder.2025}.
By \emph{relatively coarse} FE model, we mean coarse compared to the length scales of the surface topography and the relative contact displacements.
The FE mesh must still be fine enough to properly describe vibrations and wave propagation caused by the given dynamic load scenario.
The BE model relies on half-space theory, implemented on a regular grid, so that the compliance matrix can be expressed in closed form.
BE and FE model are coupled by enforcing compatibility and equilibrium conditions.
For the S4 Beam benchmark \cite{Singh.2018}, very good agreement with full-FE analysis was achieved in terms of the amplitude-dependent frequency and damping ratio of the first few modes. % higher modes were already analyzed in \cite{Linder.2025}
Compared to the full-FE analysis, the \MSM showed higher numerical robustness, even when treating rigid contact laws (without regularization) on very fine grids (ca.~25,000 nodes per contact interface), and reduced the computation effort by ca.~2-3 orders of magnitude.

\subsection{Purpose and outline of the present work}
% AIM / INTELLECTUAL MERIT AND OUTLINE
So far, the proposed \MSM has been limited to \emph{quasi-static analysis}.
The purpose of the present work is to extend the \MSM to dynamic analysis.
This is not a trivial task, since dynamic contact simulations are generally prone to numerical damping and artificial oscillations, and may require relatively small steps.
To this end, we combine the \MSM with our simulation approach for structural dynamic contact problems from \cite{MonjarazTec.2022}.
This relies on massless-boundary component mode synthesis combined with a dedicated semi-explicit time stepping algorithm.
The main scientific objectives of the present work are
\begin{itemize}
    \item to assess the computational performance of the proposed method,
    \item to ensure that the proposed method has negligible numerical damping,
    \item to numerically validate the proposed method against full-FE analysis, and
    \item to numerical explore vibration-induced settling of a bolted assembly.
\end{itemize}
%The initial intent was to simply validate this method and assess its computational performance, using full-FE simulations as reference.
%As the results will clearly show, dynamic full-FE simulations, in part, are not a suitable reference.
%This, by itself is a new and somewhat surprising finding in the field of jointed structures.
%To gain further confidence in our method, we cross-verify the time stepping results against Harmonic Balance and quasi-static analysis.
%\\
% OUTLINE
The remainder of this article is organized as follows.
The methodology, ranging from topography resolving full-FE and \MSM, to time step integration and Harmonic Balance algorithms, is described in \sref{method}.
Numerical verification, validation, and performance assessment are presented for the S4 Beam benchmark in \sref{results}.
In \sref{settling}, the vibration-induced settling behavior is explored.
Concluding remarks are given in \sref{conclusions}.

\section{Methodology\label{sec:method}}
% PROBLEM SETTING
The considered problem is the dynamics of structures subjected to frictional-unilateral interactions at contact interfaces.
The actual topography of the contact interfaces is to be resolved in sufficient detail.
We limit this work to linear behavior except for contact.
\\
We recap the full-FE method in \ssref{fullFE}, including the governing equations, and its inherent difficulties in the case of dynamic contact.
We then present the \MSM in \ssref{MSM}, including its extension towards fully-dynamic simulations.
% In the case of 3D contact, dry Coulomb friction in the tangential contact plane, and unilateral interaction in the normal contact direction are enforced.

\subsection{Full-FE analysis\label{sec:fullFE}}

\subsubsection{Problem setting\label{sec:problem}}
% GOVERNING EQUATIONS OF FULL-FE MODEL
The governing equations of the full-FE model are:
\ea{
\mm M \mm a + \mm D \mm v + \mm K \mm d &=& \mm W\lam + \fex(t) \fk \label{eq:FE}\\
\g &=& -\mm h + \mm W^{\mrm T}\mm d\fk\label{eq:cKinFE}\\
\mm p - \projC{\mm p-\epsAL\gam} &=& \mm 0 \fp \label{eq:cLawFE}
}
\erefFull{FE} describes the dynamic balance of inertia, damping, and elastic, contact and imposed forces.
Herein, $\mm d$ is the vector of nodal displacements, $\mm v = \dot{\mm d}$ and $\mm a = \dot{\mm v}$ are velocity and acceleration vector, respectively, where overdot ($\dot{\square}$) denotes derivative (of $\square$) with respect to time $t$.
Furthermore, $\mm M$, $\mm D$ and $\mm K$ are symmetric mass, viscous damping and stiffness matrices, respectively, and $\fex$ is the vector of imposed forces with known explicit time dependence.
Finally, $\lam$ is the vector of Lagrange multipliers which can be interpreted as three-dimensional contact forces, and the matrix $\mm W$ describes how the contact forces act on the FE model.
\erefFull{cKinFE} describes the contact kinematics by relating the vector of relative displacements at the integration points, $\g$, with the vector of nodal displacements $\mm d$, and the contact topography (height information of composite surface at integration points $\mm h$).
Linear contact kinematics is assumed here (small sliding), which seems reasonable as the contact interfaces remain in partial slip.
$\mm d$, $\mm h$ measure the distance from the unstressed reference configuration where the interfaces are just out of contact.
$\lam$ is sorted as $\lam = \left[\lam_1;\ldots;\lam_{\nC}\right]$, where semicolon denotes vertical concatenation, $\nC$ is the total number of integration points, and $\lam_j = \left[\lambda_{\mrm n};\lambda_{\mrm{t},1};\lambda_{\mrm{t},2}\right]_j$ contains the normal and the two orthogonal tangential components.
$\g$ is similarly sorted.
\erefFull{cLawFE} summarizes the Coulomb-Signorini conditions.
These are formulated on velocity level ($\gam=\dot{\mm g}$) and apply to all closed contacts (normal gap $g_{\mrm{n},j}=0$); \ie, \eref{cLawFE} holds only if all contacts are closed, and must be restricted to the active subset otherwise (so that the velocity of open contacts is unconstrained).
Herein, $\mm p$ is the contact stress vector, sorted analogous to $\lam$, and related to $\lam$ by spatial quadrature.
The contact conditions in \eref{cLawFE} are expressed as non-smooth equations, where $\projC{}$ denotes the projection onto the admissible set of the contact stress $\mathcal C$, and $\epsAL>\mm 0$ is a diagonal matrix.
Writing the contact constraints in this form, can be interpreted as augmented Lagrangian approach \cite{Leine.2004}.
In accordance with the sorting of $\g$ and $\mm p$, the admissible set is sorted as $\mathcal C = \mathcal C_1\times \cdots \times \mathcal C_{\nC}$, with
$\mathcal C_j = \mathbb R_0^+\times \mathcal D\left(\mu p_{\mrm{n},j}\right)$, where $\mathcal D\left(r\right)$ denotes the planar disk of radius $r$.

\subsubsection{Dynamic contact: artificial oscillations\label{sec:spurious}}
% ARTIFICIAL OSCILLATIONS; well-posedness
Conventional FE methods associate mass with each node, including those at the contact interface.
This makes the problem described by \erefs{FE}-\erefo{cLawFE} mathematically ill-posed.
An impact law has to be specified to close the equations \cite{Acary.2008}.
Regardless, the finite boundary mass leads to numerical errors:
When a contact closes (with non-zero relative speed), the velocity inevitably jumps.
In contrast to the spatially continuous case, this is linked to a finite impulse, inducing artificial high-frequency oscillations.
To stabilize the contact behavior in simulations, it is necessary to introduce numerical/algorithmic dissipation \cite{Solberg.1998,Deuflhard.2008,Bruls.2014,Acary.2013}.
As described above, this is critical when physical dissipation (friction damping) is to be analyzed.
The difficulties described occur both in the case of set-valued contact laws, and in the case of reasonably steep (penalty) regularization \cite{Tschigg.2018}.

\subsection{Multi-scale method\label{sec:MSM}}
To enable topography-resolving structural dynamic contact simulations, we propose to combine two methods, namely the \MSM \cite{Linder.2025} with the massless-boundary component mode synthesis and dedicated time step integration (\MBCMS) from \cite{MonjarazTec.2022}.
In \ssref{idea}, we recap the rationale behind the \MSM.
In \ssref{math}, we present the governing equations of the \MSM combined with the \MBCMS method.
In \ssref{TI} and \ssref{HB}, we propose time stepping and harmonic balance techniques for solving those equations.
In \ssref{algorithms}, we summarize the proposed algorithms.
In \ssref{recommendations}, we specify user recommendations, including the choice of important parameters.

\subsubsection{Rationale behind the \MSM\label{sec:idea}}
% IDEA BEHIND MSM
The \MSM proposed in \cite{Linder.2025} is illustrated in \fref{MultiScale}.
It combines the complementary strengths of the FE and BE methods.
The stress and deformation field in the contact region is modeled using elastic half-space theory, implemented on a regular and fine grid of boundary elements.
The structural dynamics of the remaining region is described using a relatively coarse FE model.
The two models are coupled by enforcing compatibility and equilibrium conditions in the far field.
\begin{figure}[t]
  \centering
  \includegraphics[width=1.0\linewidth]{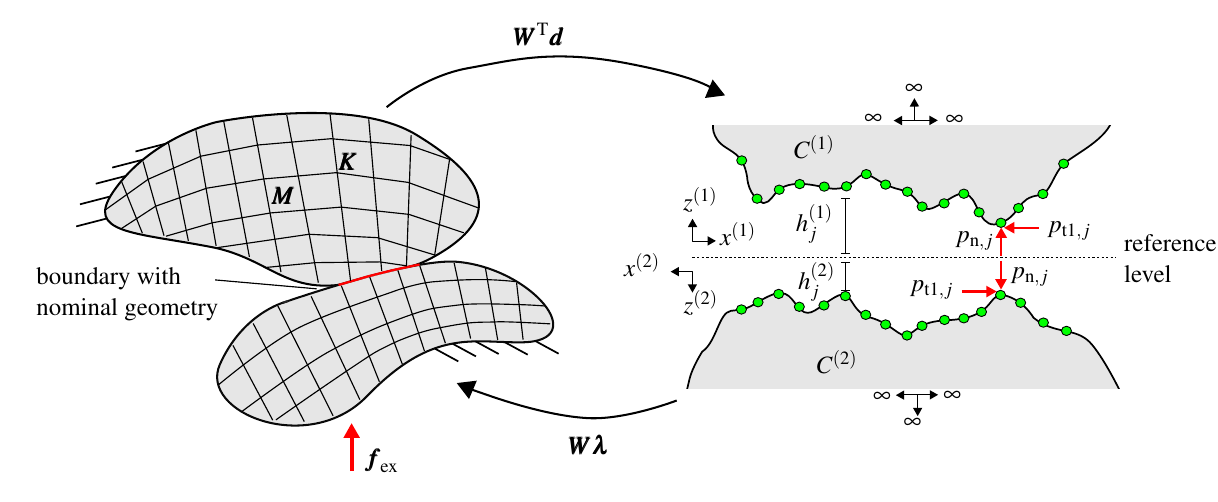}
 \caption{Schematic illustration of the proposed multi-scale method \cite{Linder.2025}.
 }
 \label{fig:MultiScale}
\end{figure}
\\
% BENEFITS
The coupled FE-BE method offers several benefits.
Most importantly, only the contact surface must be finely meshed.
The problem is further restricted to the active parts of the contact interfaces.
In contrast to FE modeling, the fine contact mesh does not propagate into the volume.
Furthermore, half-space theory in combination with a regular grid allows one to express the compliance matrix of the BE model in closed form.
Finally, the contact conditions are only imposed in the BE model, whereas the nominal contact interface in the FE model merely serves for the coupling of the two domains.
Consequently, no strict requirements are placed on the FE model at the interface.
For instance, incompatible/non-matching meshes and different element types can be quite easily handled.

\subsubsection{Mathematical formulation of the dynamic multi-scale approach\label{sec:math}}
% MOTIVATION TO DO COMPONENT MODE SYNTHESIS
As described above, the \MSM couples two domains, a relatively coarse FE model and a relatively fine BE model relying on half-space theory.
For dynamic simulations, we propose to reduce the relatively coarse FE model via \MBCMS.
Component mode synthesis is well-established for model order reduction in structural dynamics.
Specifically, we use the modified version of the Hurty-/Craig-Bampton method from \cite{MonjarazTec.2022}, which yields a singular mass matrix (with zero partitions of the mass matrix associated with the boundary).
The use of a massless-boundary model is inspired by the work of \cite{Khenous.2008}, who regard the mass associated with the contact as cause for the artificial oscillations.
\\
% MATHEMATICAL FORMULATION OF COMPONENT MODE SYNTHESIS
The modified Craig-Bampton method approximates the vector of FE nodal displacements, $\mm d$, as
\ea{
\mm d &\simeq& \matrix{cc}{
\mm I & \mm 0 \\
\tilde{\mm\Psi} & \mm\Theta
}\vector{\db\\ \di}\fk\label{eq:dT}
}
in terms of a set of Ritz vectors (called component modes; columns of the matrix on the right-hand side of \eref{dT}), and associated coefficients (generalized coordinates, stacked in vectors $\db$, $\di$).
%$\db$ corresponds to the relative displacement degrees of freedom at the nominal contact interface of the FE model.
$\db$ and $\di$ are boundary coordinates and generalized remaining coordinates, respectively.
By boundary coordinates, we mean the (relative) nodal displacements at the nominal contact interface of the FE model, and we presume conforming meshes here.
In \eref{dT}, $\mm I$ is the identity matrix, $\mm\Theta$ contains as columns the retained normal modes for fixed boundary, and $\tilde{\mm\Psi}$ contains an appropriate linear combination $\mm\Theta$ and $\mm\Psi$, where the columns of $\mm\Psi$ represent the static displacement of the remaining degrees of freedom for unit displacement of the boundary coordinates.
With this, the modified Craig-Bampton method accurately preserves the static compliance with respect to boundary loads.
Furthermore, it provides an accurate description of the dynamic compliance in the frequency range of the retained normal modes.
\\
% GOVERNING EQUATIONS OF PROPOSED METHOD
The governing equations of the \MSM combined with \MBCMS (to reduce the FE model) are:
\ea{
\kbb \db + \kbi \di &=& \wb\lam + \fexb(t) \fk \label{eq:b}\\
\mii \ai + \dii \vi + \kii \di + \kib \db &=& \fexi(t) \fk \label{eq:i}\\
\g &=& -\h +\C\lam + \wbtra \db \fk \label{eq:cKin}\\
\lam - \projC{\lam-\epsAL\gam} &=& \mm 0 \fp \label{eq:cLaw}
}
We have $\vi = \ddi$ and $\ai = \dvi$.
$\mii$, $\kii$, $\kbi$, are the respective partitions of the reduced mass and stiffness matrices.
These are obtained from the mass and stiffness matrices of the FE model by applying the \MBCMS method.
Likewise, one obtains $\fexb$ and $\fexi$ from $\fex$.
\erefFull{b} is the static force balance at the boundary of the reduced FE model, whereas \eref{i} is the dynamic internal counterpart.
\erefFull{cKin} relates the vector of relative contact displacements, $\g$, at the BE integration points to the topography (height values), $\mm h$, the elastic surface deformation of the half spaces due to the contact forces, $\mm C\lam$, and the far-field displacement of the reduced FE model, $\wbtra\db$.
As in the full-FE case, \eref{cLaw} summarizes the Coulomb-Signorini conditions.
\\
Within \eref{cKin}, $\C$ is the compliance matrix of the BE model, which is the sum of the compliance matrices $\mm C^{(1)}$ and $\mm C^{(2)}$ of the two half spaces in contact.
For a regular BE grid, the elements of $\C$ can be expressed in closed form \cite{Linder.2025}.
Also, one simply has $\mm \lambda = \mm p\Delta A$, where $\Delta A$ is the area associated with each BE integration point, and $\mm p$ is the contact stress vector.
The contact stress, as obtained from the BE model, is distributed to the boundary of the reduced FE model by the term $\wb\lam$ in \eref{b}, consistent with the principle of virtual work \cite{Linder.2025}.
$\wb$ contains the shape functions of the underlying finite elements, evaluated at the integration points of the BE model.

\subsubsection{Time step integration\label{sec:TI}}
% QUADRATURE RULES
The time stepping method in \cite{MonjarazTec.2022} relies on the leapfrog scheme with fixed time step.
The quadrature rules are
\ea{
\mm a^k &=& \frac{\mm v^{k+\frac12}-\mm v^{k-\frac12}}{\dt} \fk \label{eq:ak}\\
\mm v^k &=& \frac{\mm v^{k+\frac12}+\mm v^{k-\frac12}}{2} \fk \label{eq:vk}\\
\mm d^k &=& \mm d^{k-1}+\mm v^{k-\frac12}\dt \fp \label{eq:dk}
}
Herein, we used the abbreviation $\mm d^k=\mm d(t^k)$, $\mm v^{k+ \frac{1}{2}} = \mm v\left( t^k +  \frac{\Delta t}{2}\right)$, and so on, where $t^k$ is the $k$-th time level, and we use a fixed time step $\dt$; \ie, $t^k = t^{k-1}+\dt$.
As described above, the key idea of the time stepping scheme is (1) to explicitly step the linear internal dynamics, and (2) to implicitly treat the quasi-static contact problem.
Explicit step (1) and implicit step (2) are described in the following two paragraphs, respectively.

% EXPLICIT STEP (1)
\paragraph{Explicit step (1)}
For this step, \eref{i} is considered at time level $t^k$.
$\di^k$ and $\db^k$ are presumed given.
Restricting \erefs{ak}-\erefo{vk} to the internal coordinates, and substituting this into \eref{i}, one obtains
\ea{
\mii \frac{\vi^{k+\frac12}-\vi^{k-\frac12}}{\dt} + \dii \frac{\vi^{k+\frac12}+\vi^{k-\frac12}}{2} + \kii \di^k + \kib \db^k &=& \fexi(t^k) \fk \label{eq:itk}
}
\ea{
\Rightarrow \vi^{k+\frac12} &=& \left(\frac{1}{\dt}\mii + \frac{1}{2}\dii\right)^{-1} \left(
\fexi(t^k)
+ \left(\frac{1}{\dt}\mii - \frac{1}{2}\dii\right)\vi^{k-\frac12}
- \kii \di^k
- \kib \db^k
\right)\fp \label{eq:vikponetwo}
}
For the described \MBCMS method, $\mii$ is the identity matrix, and $\dii$ is typically diagonal.
Consequently, \eref{vikponetwo} is computationally very cheap.

% IMPLICIT STEP (2)
\paragraph{Implicit step (2)}
For this step, \eref{b} is considered at time level $t^{k+1}$.
Recall that $\db^{k}$ is presumed given, and $\di^{k+1}$ is now additionally known from the explicit step (1).
Evaluating \eref{b} at $t^{k+1}$, one obtains
\ea{
\kbb\db^{k+1} + \kbi\di^{k+1} &=& \wb\lam^{k+1} + \fexb(t^{k+1})\fk \\
\Rightarrow \db^{k+1} &=& \kbbinv\left(\wb\lam^{k+1} + \fexb(t^{k+1}) - \kbi\di^{k+1}\right)\fp \label{eq:vbkmonetwo}
}
Applying \eref{dk} to the gap, we obtain
\ea{
\mm g^{k+1} &=& \mm g^{k} + \Delta\mm g\fk\label{eq:gk}
}
where we use the auxiliary variable $\Delta\mm g = \dt \mm\gamma^{k+\frac12}$.
Substituting \erefs{vbkmonetwo}-\erefo{gk} into \eref{cKin} evaluated at $t^{k+1}$, one obtains
\ea{
\mm g^{k+1} &=& - \mm h + \C\lam^{k+1} + \wbtra\db^{k+1} \fk\\
\Delta \mm g &=& \mm G\lam^{k+1} + \mm c \fk \label{eq:Glamc}\\
\mm G &=& \C + \wbtra \kbbinv \wb \fk \label{eq:G}\\
\mm c &=& \wbtra\kbbinv\left(\fexb(t^{k+1}) - \kbi\di^{k+1}\right) - \mm g^{k} - \mm h\fp \label{eq:c}
}
Using an appropriate prediction of the gap $\mm g$, and the contact force $\lam$, one estimates the set of closed contacts.
Substituting \eref{Glamc} into \eref{cLaw}, restricted to the closed contacts, one obtains a non-smooth algebraic equation system for $\lam^{k+1}$.
The quasi-static character of the contact problem facilitates high numerical robustness and prevents artificial oscillations. %, without having to introduce algorithmic/numerical damping.
Furthermore, it allows us to use the same solver as in the quasi-static case \cite{Linder.2025}, a projected Jacobi method.
Once $\lam^{k+1}$ has been obtained, one evaluates \erefs{Glamc}, \erefo{gk}, and \erefo{vbkmonetwo}, to obtain $\Delta\mm g$, $\mm g^{k+1}$, and $\db^{k+1}$.

% STABILITY; CONVERGENCE; ETC.
The leapfrog scheme is known for its favorable energy conservation properties and second-order accuracy in the linear case.
Rigorous convergence and conditional stability theorems are available only for linear systems, and certain analytic nonlinear systems \cite{Hairer.2003}.
Such theorems do not exist in the non-smooth case, neither for the leapfrog nor for any other common scheme.
Our experience shows stable long-term integration for $\omega_{\max}\dt\lesssim 1$, where $\omega_{\max}$ is the highest retained modal frequency.

\subsubsection{Harmonic Balance\label{sec:HB}}
Besides time step integration, we implemented a Harmonic Balance analysis to directly compute time-periodic behavior.
An important motivation for this was that this allows for cross-verification:
Unlike time integration, Harmonic Balance does not suffer from numerical damping.
Thus, consistency with Harmonic Balance indicates negligible numerical damping of a given time stepping scheme.
\\
Harmonic Balance seeks periodic solutions in the form of a Fourier series,
\ea{
\db \simeq \sum\limits_{h=-H}^{H} \hat{\mm d}_{\mrm b}(h)\ee^{\ii h\Omega t}\fp \label{eq:dbFourier}
}
truncated to order $H$.
An analogous expression is used for $\di$.
In \eref{dbFourier}, $\Omega$ is the fundamental angular oscillation frequency and $\hat{\square}(h)$ denotes the $h$-th complex Fourier coefficient of quantity $\square$.
Since ${\mm d}_{\mrm b}$ is real-valued, the Fourier coefficients form complex-conjugate
pairs, $\hat{\mm d}_{\mrm{b}}(-h) = \overline{\hat{\mm d}_{\mrm b}(h)}$ for all $h\neq 0$, where $\overline{\square}$ denotes the complex conjugate.
This applies to all Fourier series considered in this work.
Harmonic Balance requires that the Fourier coefficients of the residual, obtained by substituting \eref{dbFourier} (and the analog for $\di$) into \erefs{b}-\erefo{i}, vanish up to order $H$ \cite{Krack.2019}.
This can be expressed as
\ea{
{\hat{\mm r}}_{\mrm b}(h) &=& \kbb\hat{\mm d}_{\mrm b}(h) + \kbi\hat{\mm d}_{\mrm i}(h) - \wb\hat{\mm\lambda}(h) - \hat{\mm f}_{\mrm{ex,b}}(h) = \mm 0 \quad h=-H,\ldots,H\fk\label{eq:HBb}\\
{\hat{\mm r}}_{\mrm i}(h) &=& \left(-(h\Omega)^2\mii + \ii h\Omega\dii + \kii\right)\hat{\mm d}_{\mrm i}(h) + \kib\hat{\mm d}_{\mrm b}(h) - \hat{\mm f}_{\mrm{ex,i}}(h) = \mm 0 \quad h=-H,\ldots,H\fp\label{eq:HBi}
}
The above algebraic equation system is to be solved for the sought Fourier coefficients of $\db$ and $\di$.
The problem is cast into real arithmetic and solved using a Newton-type iteration method.
\\
The iterative solution entails the computation of the Fourier coefficients of $\lam$.
To this end, an alternating frequency-time scheme is applied to the quasi-static BE model:
First, the current iterate of $\db$ is evaluated in discrete time (\eref{dbIDFT}).
Then, one successively determines the evolution of $\mm g$ and $\lam$, by quasi-statically marching from one time level to the next, as in the implicit step described in \ssref{TI}.
For periodic input $\db$, it can be shown that this yields a periodic output $\lam$ (typically after two periods).
Finally, the Fourier coefficients $\lbrace\hat{\mm\lambda}(h)\rbrace$ are determined by applying the discrete Fourier transform (\eref{lamDFT}).
\ea{
{\mm d}_{\mrm b}^k &=& \sum\limits_{h=-H}^H \hat{\mm d}_{\mrm{b}}(h)\ee^{\ii h \frac{2\pi}{N}k} \quad k=0,\ldots,2N-1\fk\label{eq:dbIDFT}\\
\hat{\mm \lambda}(h) &=& \frac{1}{N}\sum\limits_{k=N}^{2N-1}\mm\lambda^k\ee^{-\ii h\frac{2\pi}{N}k} \quad h=-H,\ldots,H\fp\label{eq:lamDFT}
}
An important benefit of the proposed approach is that the time-discontinuous quantities, $\mm g$, $\mm\lambda$ do not have to be sought explicitly in the form of a truncated Fourier series.
Instead, these quantities are implicitly treated in discrete time only.

\subsubsection{Summary of the proposed algorithms\label{sec:algorithms}}
In this section, we present three algorithms that summarize the proposed dynamic \MSM:
\emph{Algorithm 1} sets up the coupled FE-BE model.
\emph{Algorithm 2} carries out time step integration.
\emph{Algorithm 3} evaluates the residual of the Harmonic Balance analysis.
\\
As point of departure, Algorithm 1 requires the FE model.
Specifically, we need the mass and stiffness matrices, $\mm M$ and $\mm K$, and the dynamic load vector $\fex(t)$.
Also, we need the node sets at the nominal contact interfaces, the nominal node positions, the element definitions, and the degree-of-freedom map (assigning a nodal degree of freedom to each row/column of $\mm M$ and $\mm K$).
Furthermore, the topography of the composite surface is required, which we assume available as continuous function (to be able to interpolate on BE grid / FE mesh).
Finally, the area per BE grid point $\Delta A$, the highest retained modal frequency $\omega_{\max}$, and the friction coefficient $\mu$ must be specified.
Recommendations for setting those parameters are given in \ssref{recommendations}.
%%%%%%%%%%%%%%%%%%%%%%%%%%%%
\begin{algorithm}
\caption{Set up coupled FE-BE model}
\begin{algorithmic}[1]
\Require{FE model; surface topography; $\Delta A$; $\omega_{\max}$; $\mu$}
\State Reduce FE model via \MBCMS ($\mii$, $\dii$, $\kii$, $\kbi$, $\kbb$, $\fexb(t)$, $\fexi(t)$).
\State Set up BE grid.
\State Evaluate topography at BE grid points ($\mm h$).
\State Restrict BE grid to those points that may come into contact.
\State Evaluate FE shape functions at BE grid points to obtain mapping matrix ($\wb$).
\State Evaluate BE compliance matrix ($\mm C$).
\State Evaluate matrix $\mm G$ (\eref{G}).
\end{algorithmic}
\end{algorithm}
%%%%%%%%%%%%%%%%%%%%%%%%%%%%
\\
The reduction of a given FE model via \MBCMS is described in detail in \cite{MonjarazTec.2022}.
An implementation is also available in our open source repository https://github.com/maltekrack.
Specifically, the component mode synthesis method is implemented in the class \myquote{CMS\_ROM} of our tool \NLvib.
Application examples, including 3D FE models, are given in the tool \NLstep.
\\
\\
The coupled FE-BE model obtained by Algorithm 1 is the point of departure for the simulation Algorithms 2 and 3.
For time stepping (Algorithm 2), we also need initial values.
Finally, the time step $\Delta t$, the solver tolerance $\varepsilon_{\mrm{tol}}$ and the maximum number of iterations $N_{\mrm{max.iter.}}$ must be specified.
Recommendations for setting those parameters are given in \ssref{recommendations}.
%%%%%%%%%%%%%%%%%%%%%%%%%%%%
\begin{algorithm}
\caption{Time stepping of coupled FE-BE model}
\begin{algorithmic}[1]
\Require{coupled FE-BE model; initial values; $\Delta t$; $\varepsilon_{\mrm{tol}}$; $N_{\mrm{max.iter.}}$} %  solver tolerance and maximum number of iterations}
\For{$k = 1,2,\ldots$}
    \State \textbf{Explicit Step}
        \State \hskip1em $\vi^{k + \frac12} \leftarrow$ \eref{vikponetwo}.
        \State \hskip1em $\di^{k+1} = \di^{k} + \vi^{k+\frac12} \dt$.
    \State \textbf{Implicit Step}
        \State \hskip1em $\mm c \leftarrow$ \eref{c}.
        \State \hskip1em Predict gap and set of active contacts (neither sticking nor open).
        \State \hskip1em Apply projected Jacobi method to obtain $\lam^{k+1}$.
        \State \hskip1em $\Delta\mm g\leftarrow$ \eref{Glamc}, $\mm g^{k+1}\leftarrow$ \eref{gk}.
        \State \hskip1em $\db^{k+1}\leftarrow$ \eref{vbkmonetwo}.
\EndFor
\end{algorithmic}
\end{algorithm}
%%%%%%%%%%%%%%%%%%%%%%%%%%%%
\\
The projected Jacobi method relies on the successive update formula
\ea{
\lam \leftarrow \projC{\lam - \epsAL\left(\mm G\lam+\mm c\right)}\fp \label{eq:JORprox}
}
More precisely, the contact constraint enforcement is restricted to the active contacts (neither presumed sticking nor open), as described in detail in \cite{Linder.2025}.
Such a non-smooth time stepping algorithm is available also in the aforementioned public \NLstep tool.
The main difference of that implementation to Algorithm 2 is the absence of the BE model, which corresponds to the special case where $\mm C = \mm 0$ and $\wb=\mm I$ (\cf~\eref{G}).
% ALSO projective equations formulated in terms of gap velocity instead of gap increment $\Delta\mm g$
As in \NLstep, we set $\epsAL$ automatically based on the properties of $\mm G$ \cite{Studer.2009}. % Eqs. 4.59-4.60
The residuum of the non-smooth algebraic equation system is measured in terms of the L-infinity norm of the difference of $\lam$ between the last two iterations, and $\varepsilon_{\mrm{tol}}$ is the tolerance relative to the L-infinity norm of $\lam$.
\\
\\
For evaluating the Harmonic Balance residual (Algorithm 3), we need the current estimate of the Fourier coefficients of $\db$ and $\di$.
Also, the harmonic truncation order $H$, and the number of samples per period $N$ must be specified.
%%%%%%%%%%%%%%%%%%%%%%%%%%%%
%
\begin{algorithm}
\caption{Evaluate Harmonic Balance residual of coupled FE-BE model}
\begin{algorithmic}[1]
\Require{coupled FE-BE model; $H$; $N$; current estimate $\lbrace \hat{\mm d}_{\mrm b}(h)\rbrace$, $\lbrace \hat{\mm d}_{\mrm i}(h)\rbrace$; $\varepsilon_{\mrm{tol}}$; $N_{\mrm{max.iter.}}$} %  solver tolerance and maximum number of iterations}
\State ${\mm d}_{\mrm b}^k = \sum\limits_{h=-H}^H \hat{\mm d}_{\mrm b}(h)\ee^{\ii h \frac{2\pi}{N}k} \quad k=0,\ldots,2N-1$
%\State ${\mm d}_{\mrm i}^k = \sum\limits_{h=-H}^H \hat{\mm d}_{\mrm i}(h)\ee^{\ii h \frac{2\pi}{N}k} \quad k=0,\ldots,2N-1$
\State Set admissible initial values for $\mm g$, $\mm\lambda$.
\For{$k =0,1,2,\ldots,2N-1$}
        %\State \hskip1em $\mm c \leftarrow$ \eref{c}.
        %\State \hskip1em Predict gap and set of active contacts (neither sticking nor open).
        \State \hskip1em Apply projected Jacobi method to obtain $\lam^{k+1}$.
        \State \hskip1em $\mm g^{k+1}\leftarrow$ \eref{gk}. % $\Delta\mm g\leftarrow$ \eref{Glamc}
        %\State \hskip1em $\db^{k+1}\leftarrow$ \eref{vbkmonetwo}.
\EndFor
\State $\hat{\mm \lambda}(h) = \frac{1}{N}\sum\limits_{k=N}^{2N-1}\mm\lambda^k\ee^{-\ii h\frac{2\pi}{N}k} \quad h=-H,\ldots,H$
\end{algorithmic}
\end{algorithm}
%
%%%%%%%%%%%%%%%%%%%%%%%%%%%%
\\
The for loop takes care of the aforementioned quasi-static marching.
For this, one has to set initial values for $\mm g$, $\mm\lambda$.
These must be admissible in the sense that they do not violate the contact constraints.
Substituting $\Delta \mm g = \mm G{\mm\lambda}^{k+1} + \mm c$ (\eref{Glamc}) with $\mm G = \mm C$ and $\mm c = \wbtra{\mm d}_{\mrm b}^{k+1}-\mm h-{\mm g}^{k}$, into \eref{cLaw}, and recalling that $\Delta\mm g = \dt \mm\gamma^{k+\frac12}$, we obtain a non-smooth algebraic equation system for ${\mm\lambda}^{k+1}$.
This is solved with the aforementioned projected Jacobi method.
Repeating this, one successively determines the evolution of $\lam$ in discrete time.
The quasi-static marching is placed between forward and inverse discrete Fourier transforms, to map from Fourier coefficients of the coordinates to Fourier coefficients of the forces.
The entire Algorithm 3 is embedded into a numerical path continuation algorithm employing a Newton-type iteration method.

\subsubsection{Recommendations\label{sec:recommendations}}
% RANGE OF UTILITY/APPLICATION
The usefulness of the proposed method depends on the characteristics of the problem.
%       TOPOGRAPHY RELEVANT?
Resolving the \emph{actual topography} only makes sense if it is expected to play a role.
If the contacts undergo only gross slip / complete liftoff, the topography is not expected to be crucial.
In contrast, the topography is expected to be important in the partial slip/liftoff regime.
To apply the proposed method, a geometric description of the topography must be available.
%       MULTI-SCALE PROBLEM?
Furthermore, the \MSM is only useful if one actually has a \emph{multi-scale problem}.
This is the case if the FE model describing the structural dynamics can be much coarser than the BE model resolving the topography.
In other words, the shortest relevant topography wave length must be much shorter than the shortest relevant wave length of the vibrations / wave propagation.
If this is not the case, there is no motivation to use the \MSM.
Instead, one should simply use an FE model that is sufficiently fine to resolve both, structural dynamics and contact mechanics.
One should remark that the question which topography length scales are relevant for a given structural dynamics problem is still under active research.
\\
% COMPLIANCE REDUNDANCY; EDGE EFFECTS
An example where the proposed method provides no advantage is that of a smooth, complete contact (no topography).
In that case, the proposed method not only offers zero benefits, but errors are expected, which are related to the half space theory underlying the BE model.
The errors can be categorized into \emph{edge effects} and \emph{compliance redundancy}:
If the contact is concentrated near an edge of the underlying bodies, the BE model accounts for a passive elastic half space (which is not present in the actual problem setting).
Furthermore, the FE model of the underlying solid captures some compliance of the contact region already represented by the BE model.
The finer the topography is resolved, the more concentrated is the contact stress, and the less pronounced are those errors.
In practice, we recommend to verify that the ratio between closed contact area (according to the simulation) and the nominal one is small, and also to issue a warning if contact is closed near an edge.
What real-to-nominal area ratio leads to negligible compliance redundancy, depends also on the FE mesh density and the wave length of the considered vibration mode.
For the benchmarks and topographies analyzed so far, a ratio of $<10\%$ led to negligible compliance redundancy effects on the fundamental vibration modes, when using a reasonably coarse FE mesh.
\\
% USER PARAMETERS
Having clarified the recommended application range, we can now turn to the recommended setting of the parameters.
Inspecting the algorithms summarized in \ssref{algorithms}, the main user parameters are
\begin{itemize}
    \item friction coefficient $\mu$
    \item the area per BE grid point $\Delta A$
    \item the highest retained modal frequency $\omega_{\max}$
    \item the time step $\Delta t$
    \item tolerance $\varepsilon_{\mrm{tol}}$ and maximum iteration number $N_{\mrm{max.iter.}}$ of projected Jacobi method
\end{itemize}
The friction coefficient should be set according to basic tribological tests and/or experience.
The BE mesh density ($\Delta A$) must be fine enough to resolve the shortest relevant wave length of the composite surface topography.
The highest relevant frequency depends strongly on the dynamic load case.
Here, we recommend conventional estimates based on the excitation frequency spectrum and linear theory, with some margin due to nonlinear generation of higher frequencies.
Furthermore, we recommend to set the time step according to $\omega_{\max}\dt\lesssim 1$, which yields stable and converged long-term integration in our experience.
It is useful to emphasize that the recommended (and also the maximum stable) time step is \emph{mesh-independent}.
Finally, we recommend to set $\varepsilon_{\mrm{tol}}$ slightly higher than machine precision, and $N_{\mrm{max.iter.}}$ sufficiently large (\eg a few thousands).
In our experience so far, those default values never had to be adjusted.
\\
% BENEFIT: FEW PARAMETERS
Overall, it can be said that the proposed approach minimizes the number of heuristic parameters.
In particular, there is no need to specify a coefficient of restitution, nor a penalty parameter, nor any parameters related to numerical damping.

\section{Numerical results\label{sec:results}}
% PURPOSE
The aim of this section is to verify that the proposed method has negligible numerical damping, assess its computational performance, and to validate it against full-FE analysis.
%\\
% OUTLINE
\ssref{benchmark} describes the well-known S4 Beam we use as benchmark along with simulation settings.
First, we compare \MSM and full-FE method in the quasi-static case (\ssref{consistency_QSMA}).
This is a crucial intermediate step since, provided that the discrepancy is negligible in the quasi-static case, subsequently observed deviations can be attributed to time step integration and dynamic contact phenomena.
A cross-verification of the multi-scale time stepping against Harmonic Balance and quasi-static analysis is done in \ssref{consistencyModalAnalyses}.
Then, we compare the impact response obtained with the \MSM against full-FE results (\ssref{modalImpulse}).

\subsection{Benchmark system and simulation settings\label{sec:benchmark}}

\subsubsection{Geometry, material, contact and boundary conditions\label{sec:geom}}
% GEOMETRY; MATERIAL
\begin{figure}[t]
  \centering
  \includegraphics[width=1.0\linewidth]{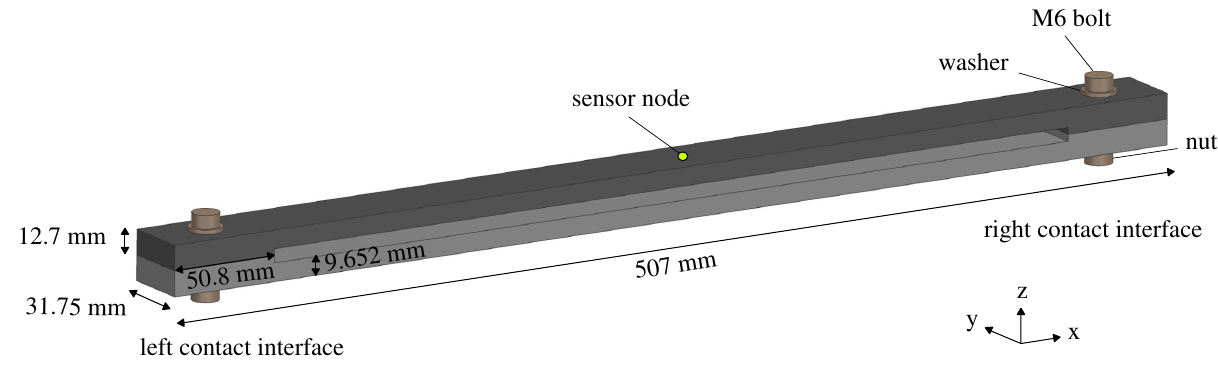}
 \caption{Considered benchmark system: S4 Beam \cite{Linder.2025}.
 }
 \label{fig:S4Beam}
\end{figure}
The S4 Beam is shown in \fref{S4Beam}.
It was originally presented in \cite{Singh.2018} and consists of two C-shaped beams which are bolted together at both ends.
The material is considered as linear-elastic and isotropic with a mass density of $\rho = 7861 ~  \mrm{kg/m^3}$, a Young's modulus of $E=194~\mrm{GPa}$ and a Poisson ratio of $\nu = 0.2854$.
\\
% TOPOGRAPHY
The topographies of the left and the right contact interface are illustrated in \fref{surfaceTopography}.
They were obtained by extracting the form deviation from surface measurements \cite{ZareEstakhraji.2023}.
The form deviation is characterized by a hill-like shape near the bore hole, flattening towards the surface edges.
The dimensions of the apparent contact area are $50.8~\mrm{mm} ~\times~31.75~\mrm{mm}$, whereby the peak-to-peak deviation due to the form is $12.14~\mrm{\mu m}$.
\begin{figure}[b]%
  \centering
  \subfloat[][]{\includegraphics[width=0.5\linewidth]{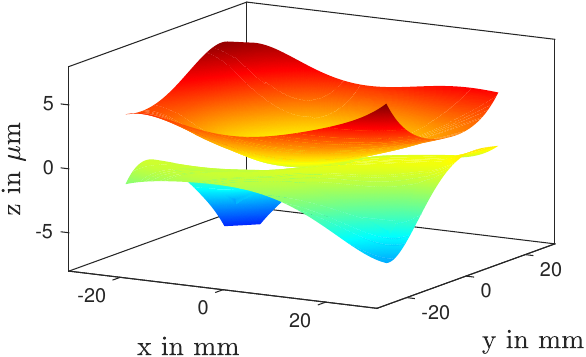}}%
  \subfloat[][]{\includegraphics[width=0.5\linewidth]{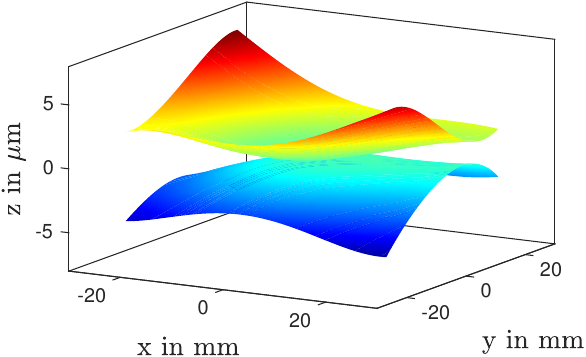}}%
  \caption{Contact topography (form deviation): (a) left, (b) right interface \cite{Linder.2025}.
  }
  \label{fig:surfaceTopography}
\end{figure}
\\
% CONTACT
Contact interactions are modeled as frictional-unilateral contact in accordance with the (set-valued) Coulomb-Signorini conditions.
The friction coefficient is set to $0.6$.
The bolts were tightened by applying a uniform area load to the nodes in the planes adjacent to the aforementioned bolt center plane. % as described in \cite{Linder.2025}.
The integral contact normal force of each interface was chosen to be $2.21~\mrm{kN}$ - similar to the experimental study in \cite{ZareEstakhraji.2023}.
\\
% BOUNDARY
Additional contact interfaces such as between bolt head and washer, between washer and beam surface, between bolt and nut, and between nut and beam surface were modeled as tied connections.
Within each bolt's center plane (orthogonal to the $z$-axis in \fref{S4Beam}), some nodes were fixed to the ambient in order to suppress rigid body modes.

\subsubsection{Mode of interest; non-frictional damping\label{sec:mode}}
% MODE OF INTEREST
For brevity, we focus on the \emph{first in-phase bending mode} in this paper.
The modal deflection shape is illustrated in \fref{Schwingform}.
\begin{figure}[b]
  \centering
  \includegraphics[width=1.0\linewidth]{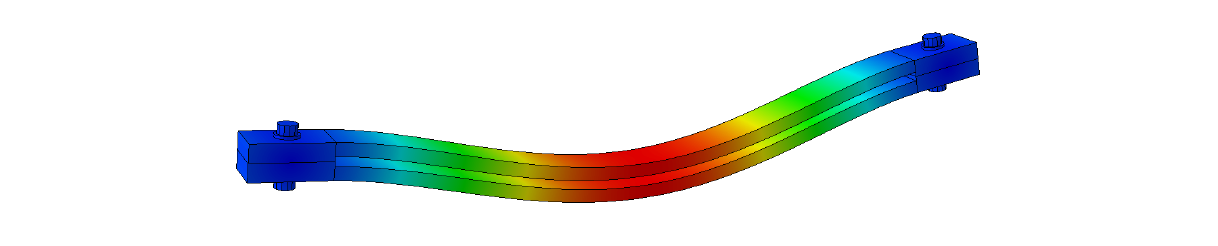}
 \caption{First in-phase bending mode shape.
 }
 \label{fig:Schwingform}
\end{figure}
This mode mainly induces partial slip at the contact interfaces in the $x$-direction (\cf~\fref{S4Beam}).
\\
% NON-FRICTIONAL DAMPING
Linear viscous damping was introduced as mass-proportional, whereby the coefficient was adjusted to obtain a damping ratio of $0.1\%$ for the first in-phase bending mode under tied contact conditions.
In a model-based prediction, one generally cannot distinguish (erroneous) numerical damping from physical damping.
Therefore, the amount of observed numerical damping should be small compared to the physical (non-frictional) damping present at low amplitudes.
It is important to emphasize that the specified linear damping ratio of $0.1\%$ is a reasonable and not at all small value, see, \eg, the low-amplitude test results in \fref{Rob}.

\subsubsection{FE and BE meshes\label{sec:FEBE}}
% DISCRETIZATION / FE and BE meshes
The central part of the beams was meshed using brick elements with quadratic shape functions, whereas linear shape functions were preferred for the beam ends including the bolts and washers.
For the full-FE analyses, a mesh was used that contains 4400 nodes at each contact interface (depicted in \fref{Meshes} (b)).
For the \MSM, a relatively coarse FE mesh with only 500 nodes per interface (\fref{Meshes} (a)) was used in combination with a regular BE grid having 4300 integration points (\fref{Meshes} (c)).
This way, full-FE analyses and \MSM have similar spatial resolution of the contact interface.
At the apparent contact interfaces, matching nodes on the opposing surfaces were enforced.
For the full-FE model, the node locations at the interface were moved in accordance with the specified topography (mesh morphing).
In the \MSM, the node locations of the considered relatively coarse FE model were not moved; instead the topography was only considered in the BE model.
\begin{figure}[t]
  \centering
  \includegraphics[width=1.0\linewidth]{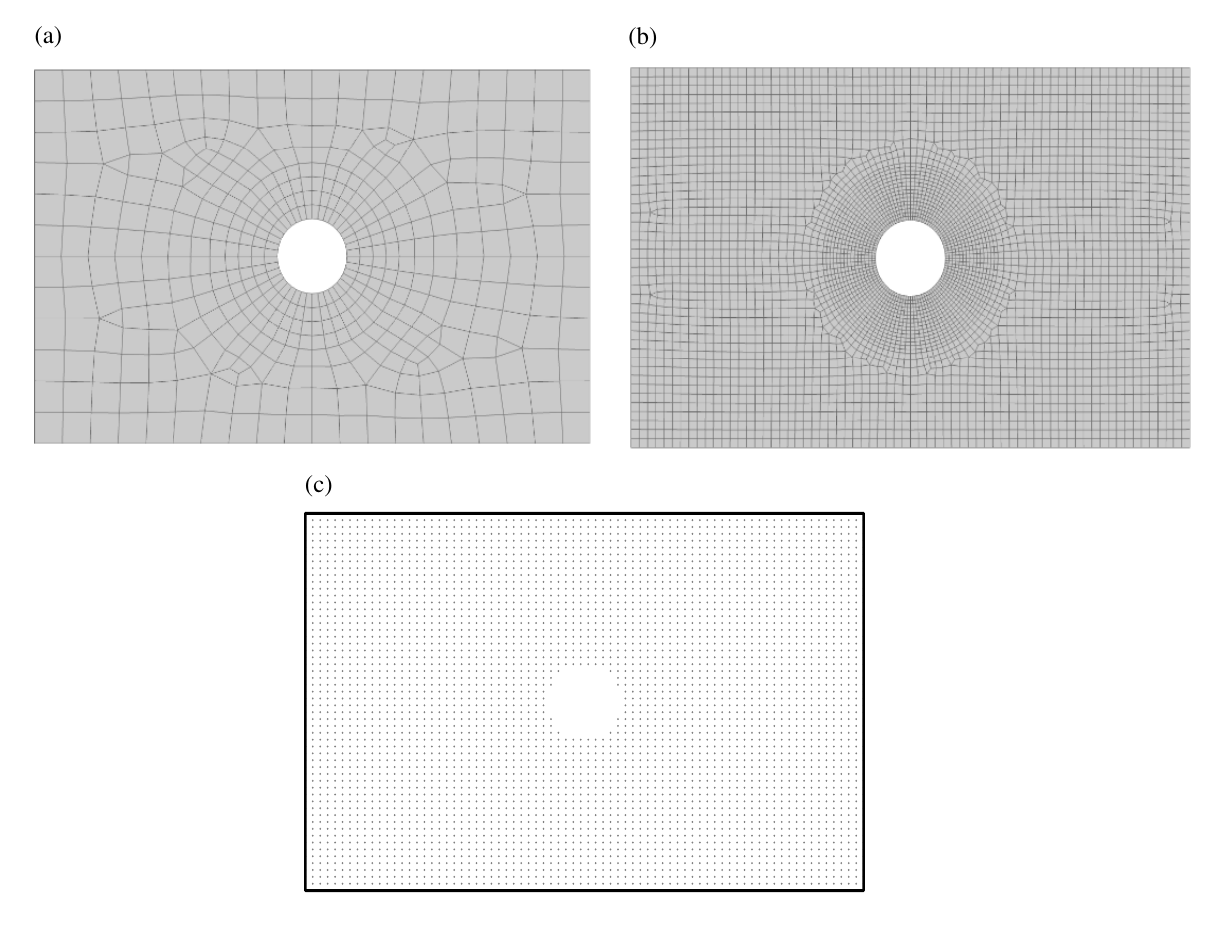}
 \caption{Interface mesh \cite{Linder.2025}: (a) mesh of the parent FE model within the \MSM, (b) contact mesh of the full-FE analysis and (c) BE grid for contact modeling within the \MSM.
 }
 \label{fig:Meshes}
\end{figure}

\subsubsection{Implementation and user parameters: \MSM\label{sec:MSMimplementation}}
The focus will be placed on an impulsive loading of half-sine wave form with a duration of $0.5~\mrm{ms}$.
The corresponding frequency spectrum quickly decays beyond $2~\mrm{kHz}$ with a secondary lobe at $4~\mrm{kHz}$.
Therefore, a value of $\omega_{\max}/(2\pi) \approx 5~\mrm{kHz}$ seems appropriate as highest retained frequency.
We supported this choice by doubling this frequency and verifying that this had negligible effect on the depicted results.
For the specified $\omega_{\max}$, we had to retain the 30 lowest-frequency normal modes within the modified Craig-Bampton method.
In addition, 3000 static modes were retained, which correspond to the 3 relative translations at the aforementioned 500 matching nodes within each of the two interfaces.
\\
The FE model, including mass and stiffness matrices, node sets, node and element definitions, and degree-of-freedom map were exported from the FE tool \ABAQUS and imported to \MATLAB.
Within \MATLAB, algorithms 1-3 (\ssref{algorithms}) were implemented, using the published versions of \NLvib and \NLstep as point of departure.
\\
To determine an appropriate time step size, we carried out a comprehensive convergence study (\aref{timeStep}).
For the \MSM, the converged time step is $\dt = 24\cdot 10^{-6}~\mrm{s}$ (\tref{computationEffort}).
With the aforementioned $\omega_{\max}/(2\pi)=5~\mrm{kHz}$, this corresponds to $\omega_{\mrm{max}}\dt = 0.74$, which is very close to our recommendation in \ssref{recommendations}.
\\
The Harmonic Balance results shown in this work were obtained with a harmonic truncation order of $H=5$, and a number of $N=50$ samples per period within the alternating frequency-time scheme.
The results did not significantly change when increasing those numbers to $H=7$ and $N=256$.

\subsubsection{Implementation and user parameters: full-FE analysis}
The reference full-FE simulations were performed using the \SIERRA tool \cite{SIERRA.2024}.
An initial attempt to complete the simulations in \ABAQUS \cite{Abaqus.2023} proved challenging due to convergence issues when attempting to model contact with direct enforcement via Lagrange multipliers.
With \SIERRA, \emph{explicit central-difference}, and an \emph{implicit HHT-alpha} scheme were used.
For the latter, the parameters were set to $\alpha = -0.1$, $\beta = 0.3025$, $\gamma = 0.6$, leading to second-order accuracy and unconditional stability in the linear case.
The explicit simulation involves mass lumping.
Large deflection effects were included, as this was observed to improve numerical robustness, even though this does not seem relevant from a physical perspective in the partial slip regime.

\subsection{Quasi-static consistency: multi-scale vs.~full-FE approach\label{sec:consistency_QSMA}}
\begin{figure}[htbp]
  \centering
  \includegraphics[width=1.0\linewidth]{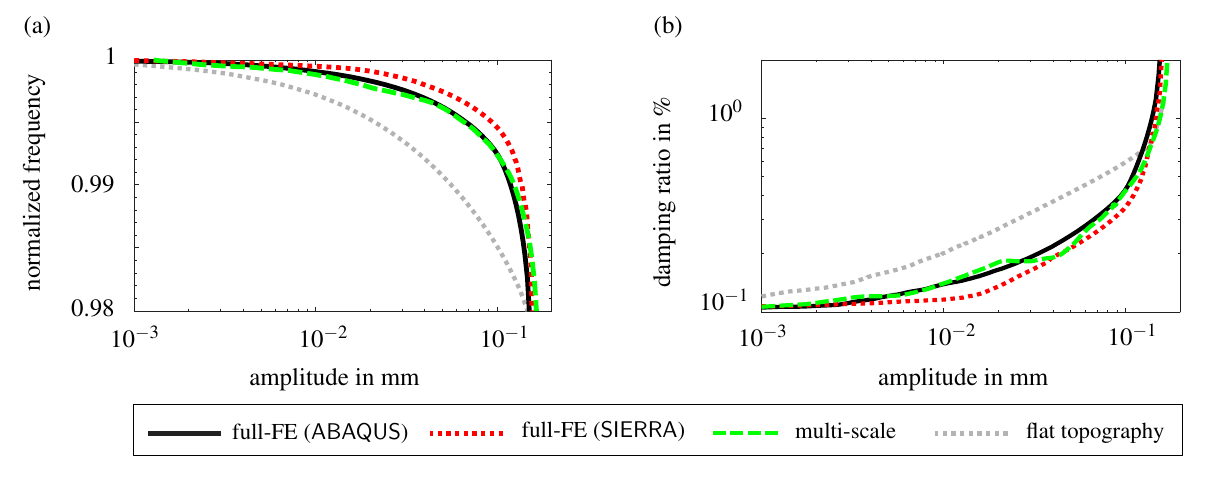}
  \caption{Amplitude-dependent properties of the first in-phase bending mode obtained with Quasi-Static Modal Analysis (QSMA): (a) frequency, (b) damping ratio. The natural frequencies are normalized w.r.t. their asymptotic values: 216.74 Hz (flat), 212.34 Hz (\ABAQUS), 209.74 Hz (\SIERRA) and 209.16 Hz (multi-scale).}
  \label{fig:modalProperties_QSMA}
\end{figure}
\begin{table}[ht]
  \centering
  \caption{\textcolor{blue}{Linear natural frequency of of the first in-phase bending mode obtained after preload step.}}
  \label{tab:lin_frequencies}
  \vspace{0.3em}
  \begin{tabular}{cccc}
    \toprule
      & full-FE (\ABAQUS) & full-FE (\SIERRA) & multi-scale\\
    \midrule
    natural frequency & 210.76 Hz & 210.09 Hz & 211.67 Hz \\
    %$\Delta \omega$ & 0.75 $\%$ & 2.87 $\%$ & 5.28 $\%$\\
    \bottomrule
  \end{tabular}
\end{table}
In this section, the consistency between the different full-FE analyses (\ABAQUS vs.~\SIERRA) and the \MSM is analyzed in terms of quasi-static analysis.
The Quasi-Static Modal Analysis (QSMA) was employed to estimate the amplitude-dependent frequency and damping ratio of the first in-phase bending mode.
QSMA was already described in \sref{intro}.
Specifically, the load $\fex = \mm M \mm\varphi_1\alpha$ is imposed in the full-FE analysis (\eref{FE}), and analogously in the multi-scale analysis (\eref{i}), where $\mm\varphi_1$ is the mass-normalized deflection shape of the first in-phase bending mode obtained after the preload step.
The load scale $\alpha$ was step-wise increased (decreased), from zero, until a maximum value $\alpha_{\max}$ (minimum value $-\alpha_{\max}$).
From the quasi-static simulation data, one can obtain an estimate of the modal frequency and damping ratio \cite{Linder.2025}.
The results are shown in \fref{modalProperties_QSMA}.
As amplitude measure, the L2 norm of the displacement at the sensor node, indicated in \fref{S4Beam}, is used.
Besides results obtained for the surface topography depicted in \fref{surfaceTopography}, for reference, full-FE results are also provided for the case of a perfectly flat surface.
\\
% FINDINGS: QUALITATIVE EVOLUTION; FLAT vs. NON-FLAT
The qualitative evolution of frequency and damping ratio with amplitude is typical for dry friction:
For sufficiently small amplitudes, the closed contact area is sticking, so that the modal parameters reach their linear limit values.
With increasing amplitude, partial slip occurs, leading to a slight frequency decrease and damping increase.
At sufficiently high amplitude, gross slip occurs, leading to a severe frequency drop, while the damping rapidly increases before it reaches a maximum, and decreases again for even larger amplitudes (not shown in \fref{modalProperties_QSMA}b).
Note that gross slip is beyond the intended technical operating range of a bolted joint.
Thus, we focus on the partial slip regime.
In that regime, there is a clear difference between ideal (flat) and actual topography.
In the flat case, the contact pressure is distributed over (almost) the entire nominal area.
Thus, the bolted region provides a higher stiffness, and the natural frequency is higher.\footnote{Throughout the figures in this paper, the modal frequency was normalized with respect to its value in the linear case, \ie, when it levels out asymptotically for small amplitudes. The respective absolute values of this frequency are given in the figure caption.}
As the outer parts of the interface see a relatively low pressure, these start sliding already at relatively small amplitudes.
Consequently, the damping ratio increases earlier in the flat case.
\\
% FINDINGS: DIFFERENCE OF TOOLS
In the non-flat case, interestingly, the full-FE results already deviate slightly in terms of the linear modal frequency (ca.~$1.2\%$ difference).
The modal frequency obtained with full-FE analysis is not converged with respect to the mesh density; \ie, an even finer mesh would be needed to reduce the linear frequency error of the full-FE model \cite{Linder.2025}.
Overall, the \MSM and the full-FE analysis yield quite consistent results.
Remarkably, the deviation between the two FE tools is larger, but still regarded as practically negligible, than the deviation between \MSM and full-FE analysis.
In this sense, the results of the \MSM are within the \emph{numerical uncertainty band} of the full-FE analysis.
From this, we conclude that the main deficiencies of the \MSM, namely edge effects and compliance redundancy, lead to negligible error here.
\begin{figure}[htbp]
 \centering
 \includegraphics[width=1.0\linewidth]{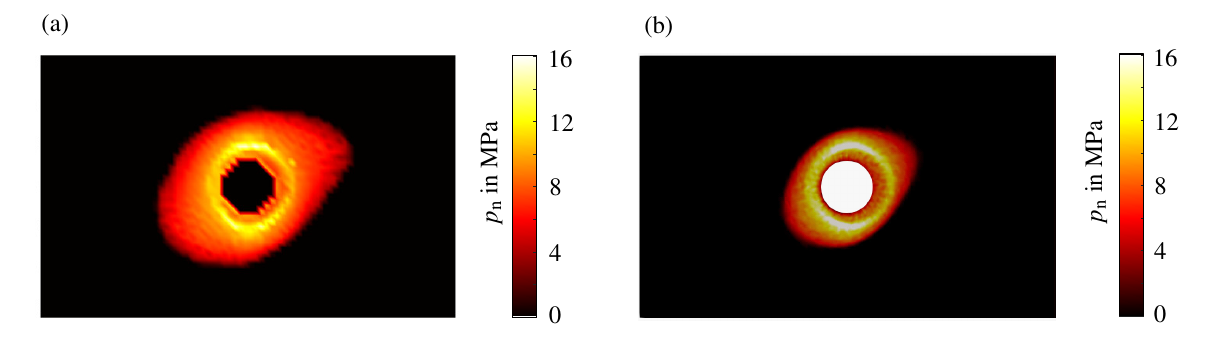}
 \caption{Contact pressure after preload in (a) for the multi-scale and (b) full-FE analysis.}
 \label{fig:preload}
\end{figure}
\\
Consistent modal parameters does not imply that the contact stress fields are identical.
Rather, potential deviations in the contact stress field might average out when projected onto the modal deflection shape.
The shorter the wave length of a mode, compared to the respective dimension of the contact interface, thus, the higher the sensitivity to local errors of the contact stress field.
In other words, identical modal parameters are a weaker criterion than identical contact stress fields.
\\
% FINDINGS based on contact pressure
The contact pressure after the preload step is shown in \fref{preload}.
As in \cite{Linder.2025}, the results are in good agreement between \MSM and full-FE analysis.
The contact pressure is distributed somewhat more uniformly, and over a slightly larger area according to the \MSM, which is attributed to the aforementioned compliance redundancy.
In addition, edge effects are visible: While the contact pressure decreases towards the bore hole in the full-FE analysis, artificial stress peaks occur there with the \MSM, since it relies on half-space theory.
\\
A detailed validation of \MSM against full-FE results in terms of contact field quantities, including stresses, is given in \cite{Linder.2025}.
That study also involves multiple topography configurations, and the analysis of multiple modes.

\subsection{Cross-verification of time stepping vs.~quasi-static analysis and Harmonic Balance\label{sec:consistencyModalAnalyses}}
% NUMERICAL DAMPING; TIME STEPPING vs. HARMONIC BALANCE (frequency response)
The purpose of this section is to check for numerical damping, which can in general be caused by time stepping in conjunction with the selected contact enforcement algorithm, of the proposed multi-scale approach.
To this end, we cross-verify the proposed multi-scale time stepping method against quasi-static analysis and Harmonic Balance, both of which do not suffer from numerical damping (associated with time stepping).
Two types of studies were done, first, a steady-state frequency response analysis under harmonic forcing, and second, nonlinear modal analysis.
%The specific analysis methods and results are described in the following.

\subsubsection{Frequency response}
For the steady-state frequency response analysis, a sinusoidal forcing was applied, $\fex = \mm M\mm\varphi_1 \alpha \cos\left(\Omega t\right)$, where the load pattern is equivalent to that used in the QSMA. % and the modal impact analysis.
The angular excitation frequency $\Omega$ was varied in the range near the primary resonance with the first in-phase bending mode.
In \fref{FRF}, results are shown for two different forcing levels, which were selected in such a way that the response level is similar to those reached after the medium- and the high-level impact considered in \ssref{modalImpulse}. %(set by parameter $\alpha$); level selected so that resonance peak ca.~at vibration level reached with medium-/high-level modal impact
The amplitude is measured in terms of the L2 norm of the displacement at the sensor node.
The amplitude-frequency relation shows a skew resonance peak, which broadens for increased excitation level.
This reflects the frequency decrease (softening) and damping increase with the amplitude, already discussed in \ssref{consistency_QSMA}, which are typical for dry friction.
\begin{figure}[htbp]
  \centering
 \includegraphics[width=1.0\linewidth]{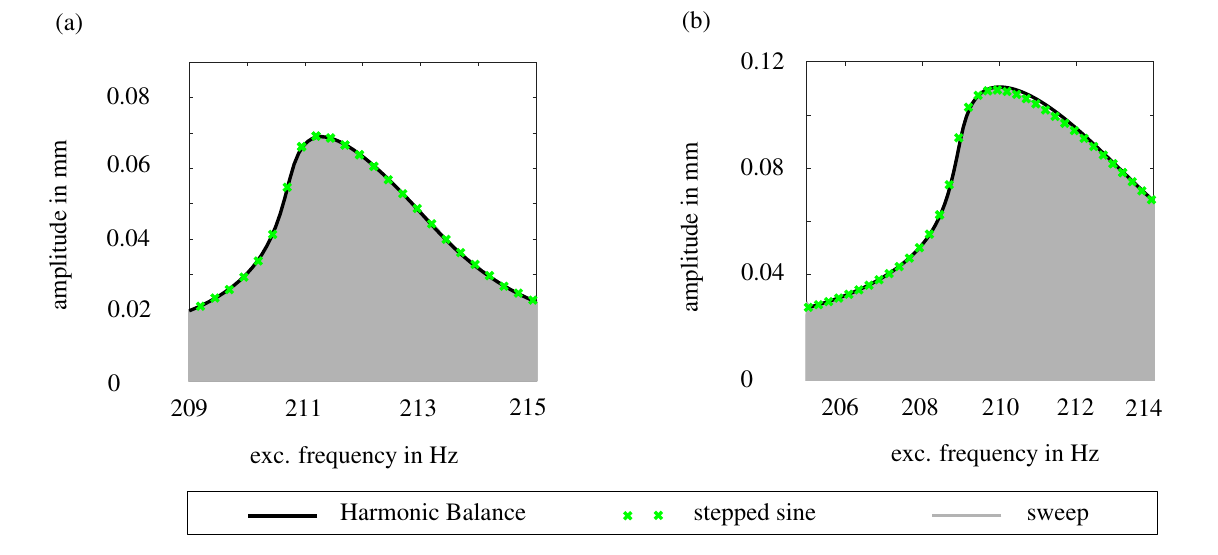}
  \caption{
  Response to harmonic modal forcing in the frequency range near the first in-phase bending mode: (a) medium, (b) high level.
  }
  \label{fig:FRF}
\end{figure}
\\
% CONSISTENCY OF TIME vs. FREQUENCY DOMAIN
With Harmonic Balance, one directly obtains the periodic steady-state response.
Recall from \ssref{comp} that the harmonic truncation order and the number of samples per period were set to $H=5$ and $N=50$, respectively.
With time step integration, a stepped frequency and a continuous frequency sweep were simulated.
For the stepped frequency simulation, $\Omega$ was fixed for $60$ excitation periods, aiming to reach a sufficiently steady state, before applying a step $\Omega\leftarrow\Omega+\Delta\Omega$, and so on.
For the frequency sweep, a fixed sweep rate was selected in such a way that a $1\%$ change in modal frequency took ca.~3500 periods, using the natural frequency/period of the linear first in-phase bending mode as reference.
From the almost perfect agreement, we conclude that there is no evidence of algorithmic/numerical damping of the multi-scale time integration method.
\\
% COMPUTATION EFFORT ANALYSIS
The simulated stepped frequency and frequency sweep tests, needed to obtain \fref{FRF}, correspond to a duration in physical time of $42~\mrm{s}$ and $23.6~\mrm{s}$, respectively.
This took $9.7$ days and $5.46$ days simulation wall time, respectively, using the computing resources specified for the multi-scale method in \tref{computationEffort}.
For reference, the simulation wall time is estimated as $2.2$ years and $1.23$ years of explicit full-FE analysis, using the data of \tref{computationEffort}.

\subsubsection{Nonlinear modal analysis}
\begin{figure}[htbp]
  \centering
  \includegraphics[width=1.0\linewidth]{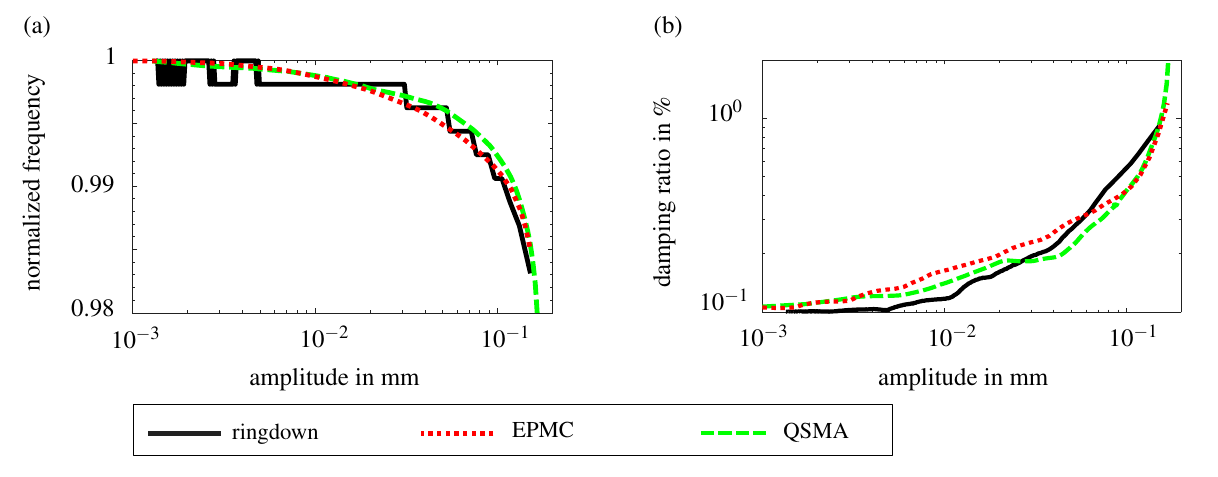}
  \caption{
  Amplitude-dependent properties of the first in-phase bending mode obtained with multi-scale approach: (a) frequency, (b) damping ratio. QSMA: Quasi-Static Modal Analysis; EPMC: Extended Periodic Motion Concept.  The natural frequencies are normalized w.r.t. their asymptotic values: 213.19 Hz (ringdown), 209.52 Hz (EPMC) and 209.16 Hz (QSMA).}
  \label{fig:modalProperties_MSM}
\end{figure}
To compute the amplitude-dependent modal frequency and damping ratio, we used three different techniques:
QSMA results are compared with those obtained via dynamic modal analysis, and the identification of modal parameters from the transient \emph{ring-down}, \ie, the free decay following an impact \cite{Kuether.2016}.
The latter two techniques are described in the following two paragraphs, respectively.
\\
% EXTENDED PERIODIC MOTION CONCEPT
For dynamic modal analysis, we used the \emph{Extended Periodic Motion Concept (EPMC)} \cite{Krack.2015}, which defines a Nonlinear Mode as a family of periodic autonomous oscillations that continuously extends a given linear mode, from the equilibrium point to finite amplitudes.
To make the autonomous oscillations periodic in the presence of (friction) damping, the EPMC relies on an artificial negative damping term, which compensates for the natural dissipation in period-average.
More specifically, the explicitly time-dependent forcing term on the right-hand side of \eref{FE} is replaced by the term $\fex = 2\zeta\omega\mm M\mm v$, where $\zeta$ is the modal damping ratio and $\omega$ is the modal angular frequency.
In the reduced problem, \eref{i}, the corresponding forcing term is $\fexi = 2\zeta\omega\mii\vi$.
In contrast to the QSMA, the EPMC consistently allows for substantial changes of the modal deflection shape, and it is well-suited for strongly nonlinear behavior and vibration-induced normal load variation (and opening-closing events).
To compute the periodic modal oscillations, we used Harmonic Balance.
Truncation order and number of samples per period were set as in the frequency response case.
\\
% MODAL PARAMETER IDENTIFICATION FROM RING-DOWN DATA
For the ring-down analysis, a \emph{modal impact} was applied in the form of a half-sine,
\ea{
\fex = \begin{cases}
    \mm M \mm\varphi_1\sin\left(\pi \frac{t}{T}\right) & 0\leq t\leq T \\
    \mm 0 & t\geq T
\end{cases}\fk \label{eq:halfsine}
}
of duration $T=0.0005~\mrm{s}$.
The forcing term is analogous in the reduced problem (\eref{i}).
Note that the load pattern, $\mm M \mm\varphi_1$, ensures that only the fundamental mode responds in the linear case.
The transient response was simulated using time step integration.
Subsequently, the short-time Fourier transform was applied in conjunction with a Hanning window.
This yields an estimate of the instantaneous frequency and amplitude.
The instantaneous damping ratio was estimated from the decay of the fundamental harmonic response \cite{Kuether.2016}.
\\
% FINDINGS
The results are shown in \fref{modalProperties_MSM}.
Recall that all data was obtained from the same multi-scale model.
Overall, the modal analysis techniques are in very good agreement.
The fact that the QSMA results are well-aligned with the EPMC results implies that the simplifying assumptions underlying QSMA are valid.
More specifically, the modal deflection shape is in good approximation amplitude-constant, and the contact behavior is dominated by friction without substantial normal load variation.
This insignificant normal load variation is expected since the considered in-phase bending mode mainly induces slip, and the negligible mode shape variation is expected within the partial slip regime.
The ring-down results are well-aligned with QSMA and EPMC in terms of damping, while there is a slight offset in frequency of ca.~$2\%$, which is explained by settling later (\ssref{settling}).
The consistency in terms of damping implies that the impact response contains negligible secondary modal contribution.
This is expected for a modal impact in the weakly-nonlinear regime.
It is important to note that the modal damping ratio levels out at $0.1\%$, \ie, the value specified as linear viscous damping.
Together with the results of the Harmonic Balance analysis, we summarize that the proposed multi-scale time integration methods shows no evidence of numerical/algorithmic damping, neither in the linear nor in the nonlinear case.
It is also noteworthy that the consistency of QSMA and ring-down results is rarely analyzed, especially for topography-resolving models.

\subsection{Consistency of modal impact response\label{sec:modalImpulse}}
\begin{figure}[htbp]
  \centering
  \includegraphics[width=1.0\linewidth]{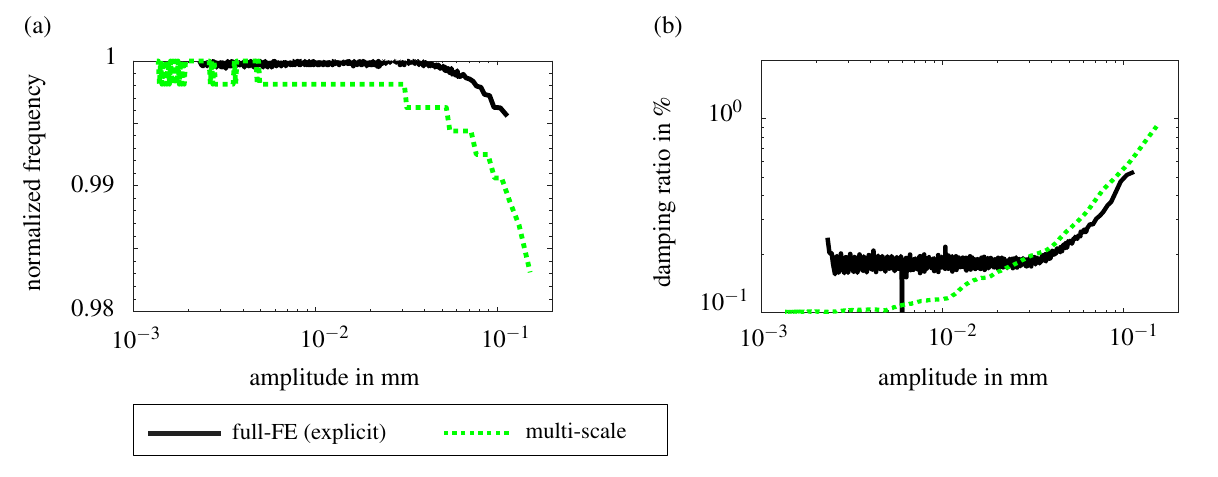}
  \caption{Amplitude-dependent properties of the first in-phase bending mode obtained from modal impact response: (a) frequency, (b) damping ratio. The natural frequencies are normalized w.r.t. their asymptotic values: 213.19 Hz (multi-scale) and 210.37 Hz (full-FE).}
  \label{fig:modalProperties_ringdown}
\end{figure}
In this section, we compare results obtained with the \MSM to full-FE results.
\fref{modalProperties_ringdown} shows the amplitude-dependent frequency and damping ratio of the first in-phase bending mode estimated from ring-down data.
Modal impact loading and modal parameter identification were specified and done as before (\ssref{consistencyModalAnalyses}).
Remarkably, the modal damping ratio obtained with explicit full-FE analysis levels out at almost twice the specified linear value of $0.1\%$.
We attribute this to numerical damping.
This was also made responsible for the over-prediction of the damping in the TRChallenge (\cf Fig.~15-right in \cite{Krack.2025}).
\begin{figure}[htbp]
  \centering
   \includegraphics[width=1.0\linewidth]{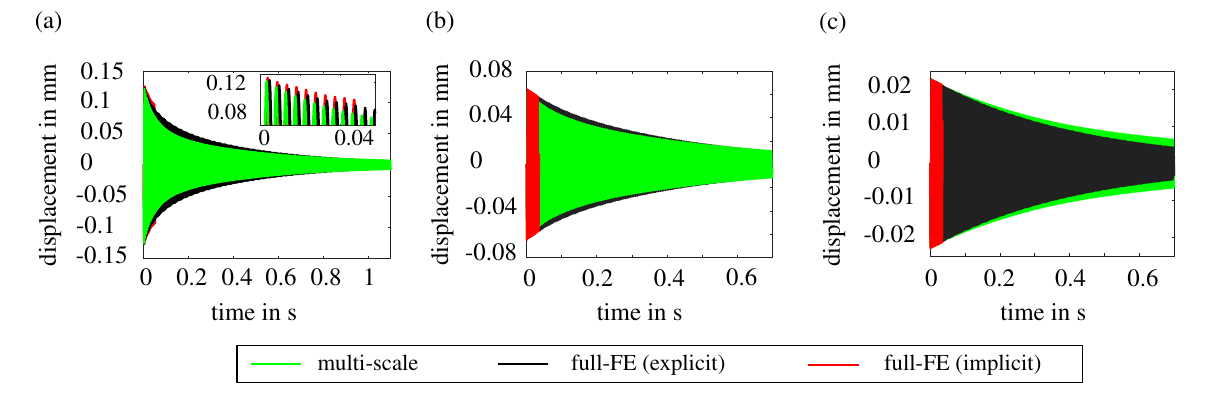}
  \caption{Time evolution of sensor node $z$-displacement for modal impact of (a) high, (b) medium and (c) low level.
  }
  \label{fig:ringdown}
\end{figure}
\\
% FINDINGS based on time evolution
\fref{ringdown} shows the time evolution of the sensor node $z$-displacement for three different impact strengths.
In full accordance with \fref{modalProperties_ringdown}, the \MSM results initially decay faster, whereas it is the opposite at lower amplitudes.
Recall that the faster decay of the explicit full-FE simulation data, at lower amplitudes, is attributed to numerical damping.
At high amplitudes, there is a clear deviation between explicit and implicit full-FE simulations.
The results suggest that the explicit scheme has higher numerical damping, also in the nonlinear regime.

\subsection{Computational performance\label{sec:comp}}
Throughout all studies we have done so far, we observed excellent numerical robustness of the proposed \MSM.
For the given benchmark system, the projected Jacobi method usually required between 10 and 100 iterations.
This seems relatively high.
Therefore, we implemented a semi-smooth Newton method with analytical gradients.
With this, the number of iterations is usually much less than 10.
However, the cost per iteration is much larger, since the gradients must be evaluated and factorized.
In contrast, the projected Jacobi method is gradient-free.
Overall, the projected Jacobi method led to lower computation effort according to our observations so far.
\\
An overview of the computational effort of \MSM and full-FE analysis is given in \tref{computationEffort}.
The selected time steps are the result of a thorough convergence study, which is described in \aref{timeStep}.
\\
An important motivation to use implicit time step integration methods is that they allow larger time steps.
Interestingly, this is not the case here for the full-FE method.
While larger time steps might have been possible without instability, a relatively small time step is needed to achieve converged results.
And this time step is not larger than the converged time step of the explicit method.
\\
The \MSM permits a larger time step than the full-FE method.
It is useful to recall that the largest stable time step of the full-FE method is restricted by the mesh size (and the material properties, in accordance with the Courant-Friedrichs-Lewy condition).
In contrast, the largest stable time step is mesh-independent in the \MSM, where it is instead bounded by the highest retained frequency.
The highest retained frequency, in turn, depends on the dynamic loading and the resulting structural dynamics (\cf~paragraph 1 in \ssref{MSMimplementation}).
In contrast, the mesh size in the full-FE method is usually restricted by the need to resolve the contact topography in a sufficiently fine way.
\\
The simulation wall time generally depends on the available computing platform, and to what extent the implementation is optimized for the given platform.
\SIERRA is optimized for high-performance computing and ran on an appropriate cluster (\cf title of \tref{computationEffort}).
In contrast, our ad-hoc \MATLAB implementation ran on a low-/medium-performance machine.
This makes sure that the computational speedup is biased in favor of the full-FE simulation.
Hence, the reported simulation speedup by a factor of ca.~80 and 400, of the \MSM compared to explicit and implicit full-FE analysis, is to be viewed as a lower bound.
However, for a given simulation task, the computational overhead of the \MSM should be additionally accounted for.
This consists of setting up the coupled FE-BE model (Algorithm 1 in \ssref{algorithms}), which took ca.~200 s wall time.
\begin{table}[t]
\caption{Computational effort. The full-FE simulations ran on a high-performance cluster (Intel Broadwell E5-2695 machine with 36 threads). The multi-scale simulations ran on a well-equipped personal computer (Intel i9-13900K machine with 24 CPUs).}
\begin{center}\label{tab:computationEffort}
\begin{tabular}{l l l}
\toprule
\rule[-1ex]{0pt}{2.5ex}
Metric & multi-scale &  full-FE \\
\midrule
\midrule
\rule[-1ex]{0pt}{2.5ex} nodes/points in contact mesh/grid & 4300 & 4400 \\
& (26\% in potential contact) & \\
\hline
\rule[-1ex]{0pt}{2.5ex} wall time for preload step & 74 s & 17.5 min \\
\hline
\rule[-1ex]{0pt}{2.5ex}  wall time for quasi-static modal analysis & 107 s & 5.3 h \\
\hline
\rule[-1ex]{0pt}{2.5ex} wall time for simulating 1 ms ring down & & \\
\rule[-1ex]{0pt}{2.5ex} explicit & 20 s & 27.5 min \\
\rule[-1ex]{0pt}{2.5ex} implicit & - & 2 h 23 min \\
\hline
\rule[-1ex]{0pt}{2.5ex} converged time step & & \\
\rule[-1ex]{0pt}{2.5ex} explicit & $24\cdot 10^{-6}~\mrm s$ & $12.5 \cdot 10^{-6}~\mrm s$ \\
\rule[-1ex]{0pt}{2.5ex} implicit & - &  $6.25\cdot 10^{-6}~\mrm s$ \\
\hline
\end{tabular}
\end{center}
\end{table}

\section{Settling\label{sec:settling}}
This section is devoted to the particularity of frictional systems that they may ring down to different static equilibrium configurations.
It is useful to briefly recap the specific state of knowledge (\ssref{stateOfKnowledgeSettling}), including the classical theorems and engineering relevance.
Subsequently, we show evidence of the described settling behavior obtained from both, \MSM and full-FE method (\ssref{evidence}).
To the best of our knowledge, this is the first study that shows numerical evidence of settling of a bolted assembly due to structural vibrations.
Finally, we demonstrate that the \MSM is well-suited to explore this interesting phenomenon, among other interesting studies not included here.

\subsection{State of knowledge\label{sec:stateOfKnowledgeSettling}}
% COULOMB --> SET-VALUED EQUILIBRIUM; MEMORY
The friction stress of a sticking point can assume an arbitrary value within the limits given by the Coulomb law (inside of Coulomb disk).
According to the balance of momentum, each admissible friction stress distribution corresponds to a slightly different static equilibrium configuration.
The relative tangential displacement field which remains after the vibrations have decayed is referred to as residual slip field.
The residual slip depends on the entire load history, and is generally unknown.
In this sense, permanently sticking parts of a contact interface possess memory.
Residual slip, associated friction stress and static equilibrium configuration are generally \emph{set-valued}.
\\
% KLARBRING THEOREM; COUPLING
When all contacts experience recurrent sliding and/or lift-off (no permanently sticking contacts), the system eventually loses its memory; \ie, the long-term behavior is independent of the initial static equilibrium \cite{Klarbring.2007}.
Conversely, if at least one contact is permanently sticking, the long-term behavior generally depends on the residual slip in those contacts.
In particular, this is the case if the contacts are \emph{coupled}, meaning that the normal displacement is influenced by the friction force and vice versa.
Such a coupling may stem from the pairing of dissimilar materials \cite{Barber.2008,Ponter.2016,Flicek.2017}, the alignment of two contacts under an angle (geometric coupling), or the elastic coupling between normal and tangential contact directions via the underlying flexible structures, not necessarily with respect to the same contact area, but possibly among different contact areas (\eg, the tangential displacement of one contact area affects the normal stress at another).
\\
% FRICTION DAMPERS
The variability of the vibration behavior due to the set-valued residual slip is known to be of high technical relevance for friction dampers.
Experimental results show that the resonant vibration level a system with friction dampers can vary by as much as $25\%$ or even $300\%$ depending on the load history \cite{Botto.2018,Gastaldi.2021}.
\\
% JOINTED STRUCTURES
In the realm of jointed structures, \ie, assemblies connected via fasteners like bolts or rivets, not much is known about the effect of settling (or set-valued residual slip) on the vibration behavior.
Settling is viewed as one of the main reasons for non-repeatability of vibration tests involving jointed structures \cite{Brake.2021}.
It is common practice to run repeated vibration tests until the results stabilize, and to report only those final measurements (\cf~\fref{Rob}).
Measurements obtained during vibration-induced settling are rarely reported \cite{Bhattu.2025}.
It follows from the above theory that the variability approaches zero when transitioning towards the gross slip / complete lift-off regime.
This is also observed in experiments of jointed structures, see \eg Fig.~10 of \cite{Roettgen.2017} where a catalytic converter system was studied.
\\
% ANALYSIS METHODS
The theoretical investigation of settling of jointed structures is hampered by the lack of analysis methods.
QSMA uses the static equilibrium after the preload step as point of departure, and is therefore unable to account for settling.
Recently, Harmonic-Balance-based modal analysis has been extended to  account for residual-slip-related variability \cite{Ferhatoglu.2026}.
So far, however, this approach is restricted to a small number of contacts / very coarsely resolved contact interfaces.
Transient simulation (by time step integration), has the important benefit that it allows to directly analyze the effect of the load history (initial assembly; quasi-static and dynamic loads) on the residual slip.

\subsection{Evidence\label{sec:evidence}}
% FINDINGS based on slip ratio distribution
In \fref{slipratio}, the slip ratio is depicted, \ie, the norm of the friction stress divided by friction coefficient times contact pressure (\cf~\fref{preload}).
The slip ratio was evaluated after the vibrations have decayed, following the high-level impact shown in \fref{ringdown}a.
We refer to this as \emph{residual slip ratio} as opposed to the initial slip ratio caused by the preload step.
The residual slip ratio depends on the strength of the applied impact, as shown in \ssref{explore}.
\MSM and full-FE results are in good qualitative and reasonable quantitative agreement.
The settling process is expected to be sensitive to the parameters of the numerical simulation algorithm, and this explains to some extent the visible deviations between \MSM and full-FE analysis.
\begin{figure}[htbp]
 \centering
 \includegraphics[width=1.0\linewidth]{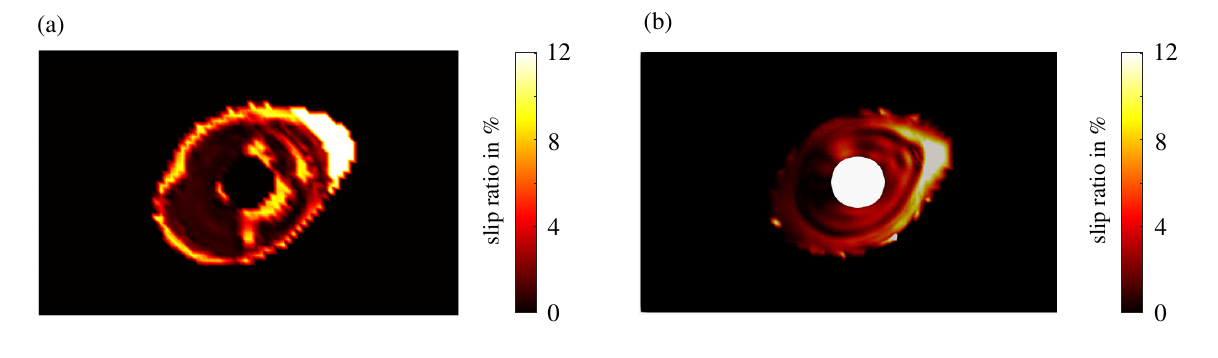}
 \caption{Slip ratio after decay an impact event: (a) multi-scale analysis and (b) full-FE analysis. The results correspond to the ring down from the high-level impact shown in \fref{ringdown}a.}
 \label{fig:slipratio}
\end{figure}
\\
% SETTLING
The results in \fref{slipratio}, obtained with both full-FE and multi-scale simulations, are clear evidence of \emph{vibration-induced settling}.
The system rings down to a static equilibrium that differs slightly from that after the preload step, \ie, before a dynamic load is applied.

\subsection{Exploration using the \MSM\label{sec:explore}}
\begin{figure}[htbp]
  \centering
  \includegraphics[width=1.0\linewidth]{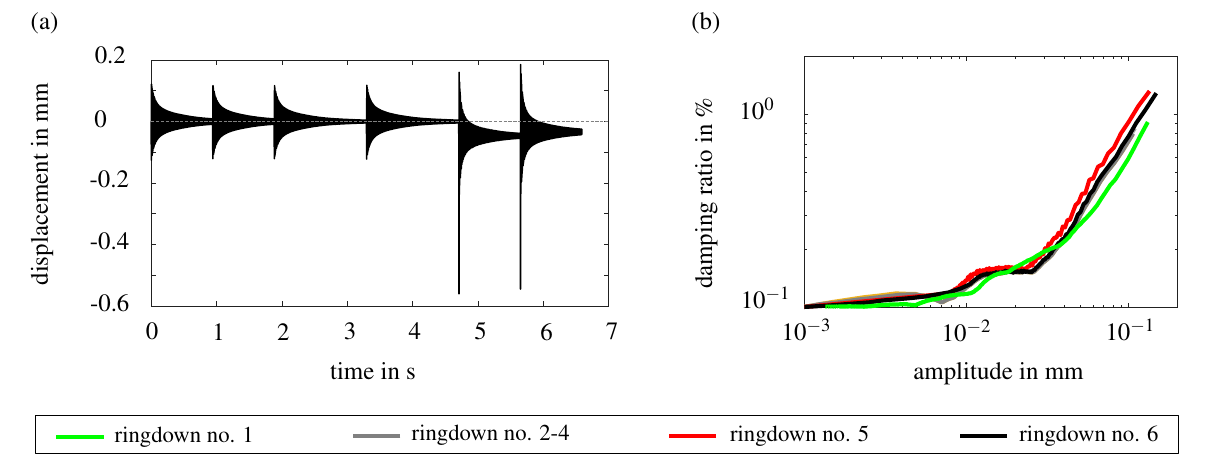}
  \caption{Settling along a modal impact sequence: (a) evolution of sensor node $z$-displacement; (b) evolution of amplitude-dependent modal damping ratio.}
  \label{fig:multiRingdown}
\end{figure}
\begin{figure}[htbp]
  \centering
  \includegraphics[width=1.0\linewidth]{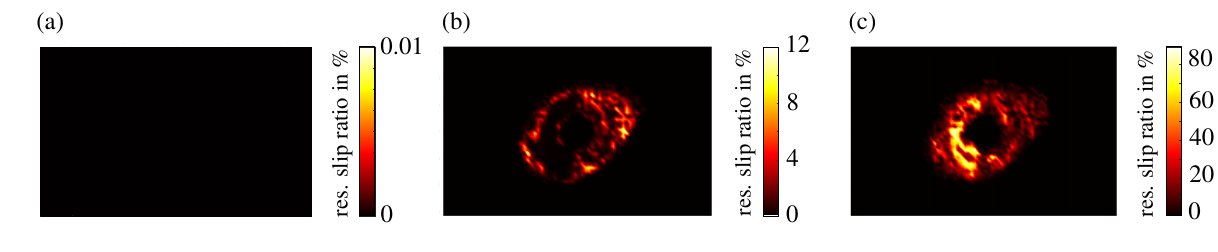}
  \caption{Residual slip ratio: (a) after preload; (b) after second impact; (c) after fifth impact.}
  \label{fig:residualTractions}
\end{figure}
In this section, a sequence of impacts is considered, and the resulting evolution of the static equilibrium is analyzed.
The results in this section are limited to the \MSM.
The considered scenario spans ca.~$7~\mrm{s}$ in physical time, which took less than 2 days with the \MSM.
For the explicit full-FE simulation, we estimate a wall time of more than 130 days, using the data from \tref{computationEffort}.
\\
A sequence of 6 impacts is applied.
The first impact has the same form and strength (high-level) as that considered in \fref{modalProperties_ringdown} (and also in \fref{ringdown}a), leading to a maximum displacement at the sensor node of ca.~$0.13~\mrm{mm}$.
Subsequently, an impact of the same strength is applied after $1~\mrm{s}$, another $1~\mrm{s}$, and another $1.4~\mrm{s}$, respectively.
Finally, an impact of four times that strength (super-high-level) is applied after another $1.4~\mrm{s}$, and another $1~\mrm{s}$, respectively.
The resulting evolution of the displacement at the sensor node is depicted in \fref{multiRingdown}a.
\\
At approximately $0.3~\mathrm{s}$ after each impact, nearly the entire contact interface is stuck (with the exception of a few peripheral spots); \ie, the contact settles well before the next impact is applied.
The residual slip ratio is shown, after the preload, the second (high-level) impact, and the fifth (super-high-level) impact in \fref{residualTractions}.
To determine the residual state in terms of contact stress and displacement, the time-average is taken over the last $10$ vibration periods preceding the subsequent impact.
%\\
As already mentioned at the end of \ssref{consistencyModalAnalyses}, the residual traction obtained after even a low-level impact is non-negligible, in contrast to the initial residual traction obtained after the preload (\fref{residualTractions}).
During the first high-level impact, maximum relative sliding distances of $1.69~\mum$ were observed, while the residual slip measured up to $0.25~\mum$. %, and $0.05~\mum$ at the sensor node.
\\
The super-high-level impact is sufficient to briefly induce gross slip at the contact interfaces.
Here, maximum relative sliding distances of $43.8~\mum$ were observed, while the residual slip measured up to $8.79~\mum$. %remained in the contact, and $0.046~\mathrm{mm}$ at the sensor node.
Consequently, the residual slip ratio is quite different and larger in magnitude than that obtained after the previous impacts.
The associated change of the static equilibrium position is also visible in \fref{ringdown}a at the sensor node, \ie, far away from the contact interfaces, where it amounts to ca.~$50~\mum$.
It seems feasible but challenging to acquire such a change in static equilibrium position by appropriate instrumentation in an experiment.
% 0.8 MPa vs. 7 MPa maximum friction stress
\\
As expected, the contact settles upon repeated impacts of the same strength.
Thus, the ring-down data obtained after impacts 2-4 yields the same modal parameter results (\fref{multiRingdown}b).
One should remark that the effect of settling on the amplitude-dependent modal damping in \fref{multiRingdown}b is relatively small.
We conjecture that the effect would be larger if misalignment of the contact interfaces was accounted for.

\section{Conclusions\label{sec:conclusions}}
The proposed multi-scale time stepping method showed excellent computational performance.
Specifically, it demonstrated high numerical robustness and no evidence of numerical/algorithmic damping, as cross-verified by quasi-static and Harmonic Balance results.
Compared to full-FE simulations, the wall time was reduced by several orders of magnitude.
We established practical recommendations on the target application range and the setting of the user parameters.
In particular, the proposed approach enables analysis of the effect of the actual contact topography on the dynamics of jointed structures.
\\
For the given problem class of topography-resolving dynamic contact simulations, FE analysis produced a spread of results, in spite of thorough convergence studies.
This suggests that full-FE simulations cannot always serve as absolute reference for the given problem class.
In particular, we found errors related to numerical damping in the same order as typical structural damping.
In addition, the computation is either infeasible or the effort is too high to be practically useful, even for the simple S4 Beam benchmark with just two bolted joints.
\\
In lack of a suitable numerical reference, the next step is to use experimental data for validating the proposed simulation method.
Furthermore, we aim to mitigate the limitations of the multi-scale method, including edge effects and compliance redundancy.
Finally, we wish to use the proposed method to address fundamental research questions related to, \eg, vibration-induced settling, impact-induced mode mixing, and the effect of different length scales of the composite contact interface topography on the dynamics.

%%%%%%%%%%%%%%%%%%%%%%%%%%%%%%%%%%%%%%%%%%%%%%%%%%%%%%%%%%%%%%%%

\section*{Acknowledgements}
We are grateful for the funding received by the Ministry of Science, Research and Arts Baden-Württemberg, Germany (FKZ Land: MWK32-7531-49/13/7), and the Deutsche Forschungsgemeinschaft (DFG, German Research Foundation) [Project 450056469].

\appendix

\section{Time step convergence study\label{asec:timeStep}}
\begin{figure}[htbp]
  \centering
  \includegraphics[width=1.0\linewidth]{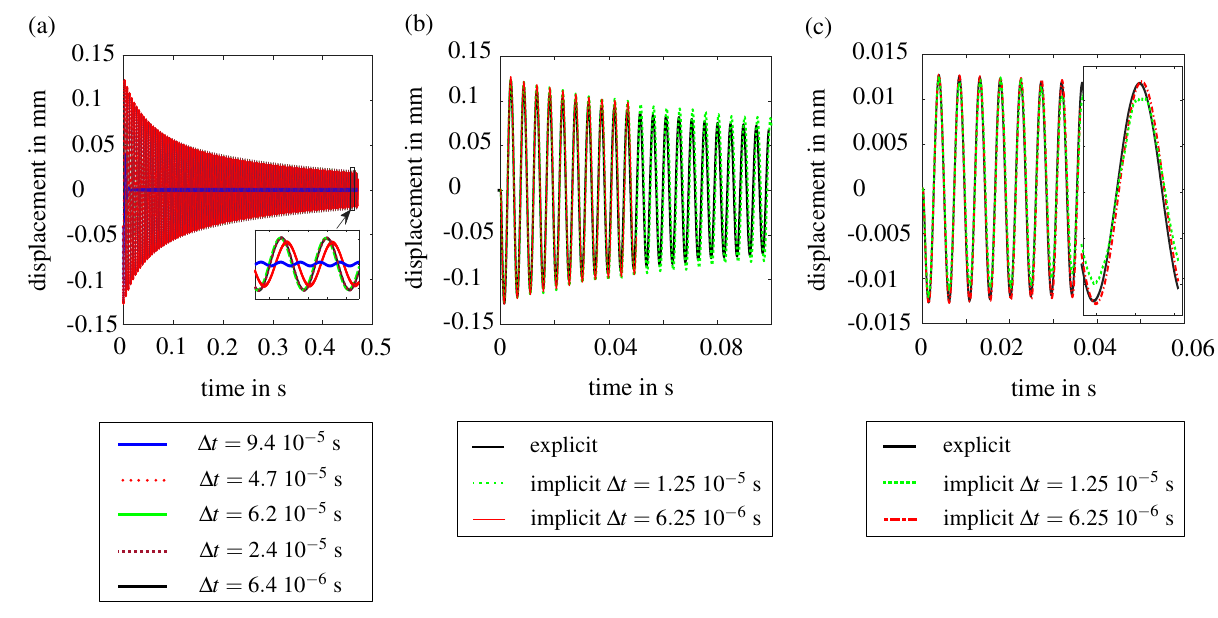}
  \caption{
  Time evolution of sensor node $z$-displacement after modal impact of (a)-(b) high and (c) low level, for different time step sizes.
  (Left) multi-scale, (center) and (right) full-FE simulation.
  }
  \label{fig:timeStep}
\end{figure}
To gain confidence in the implicit and explicit full-FE as well as the multi-scale simulations, we carried out a thorough time step convergence study.
Representative results are shown in \fref{timeStep}.
As load case, a modal impact (\cf~\eref{halfsine}) of high / low strength is considered, and the time evolution of the sensor node $z$-displacement is depicted, for different time step sizes $\dt$.
It can be verified that the results obtained with the different methods are stabilized for the respectively selected time step size (\cf~\tref{computationEffort}).
Again, it should be noted that the stabilized results differ between implicit and explicit full-FE simulations, where the explicit simulation exhibits higher numerical damping.

\end{document}